\documentclass[fleqn,usenatbib]{mnras}
\usepackage{newtxtext,newtxmath}
\usepackage[T1]{fontenc}
\DeclareRobustCommand{\VAN}[3]{#2}
\let\VANthebibliography\thebibliography
\def\thebibliography{\DeclareRobustCommand{\VAN}[3]{##3}\VANthebibliography}

\usepackage{graphicx}	
\usepackage{subcaption}
\usepackage{amsmath}	
\usepackage{comment}

\usepackage{multirow}
\usepackage{ragged2e}
\usepackage{longtable}
\usepackage{CJKutf8}
\usepackage{subcaption}
\usepackage{rotate}
\usepackage{pdflscape}
\usepackage{float}
\usepackage{xcolor}
\usepackage{placeins}

\title[Characterisation of all known multiple stellar systems within 10\,pc]{Characterisation of all known multiple stellar systems within 10\,pc}

\author[J. González-Payo et al.]{
J. González-Payo,$^{1}$\thanks{E-mail: {\tt fcojgonz@ucm.es}}
J.\,A. Caballero,$^{2}$
C. Cifuentes,$^{2}$
M. Cortés-Contreras,$^{1}$
and F. Rica$^{3}$
\\
$^{1}$Departamento de Física de la Tierra y Astrofísica \& IPARCOS-UCM (Instituto de Física de Partículas y del Cosmos de la UCM),\\ Facultad de Ciencias Físicas, Universidad Complutense de Madrid, 28040 Madrid, Spain\\
$^{2}$Centro de Astrobiología (CSIC-INTA), European Space Astronomy Centre, Camino Bajo del Castillo s/n, 28692 Villanueva de la Cañada, Madrid, Spain\\
$^{3}$Federaci\'on Extreme\~na de Astronom\'ia, c/ Jos\'e Ruiz Azor\'in 14, 4$^\circ$~D, 06800 M\'erida, Spain
}

\date{Accepted 2026 April 28. Received 2026 April 28; in original form 2026 February 24}

\pubyear{\the\year{}}

\begin{document}
\label{firstpage}
\pagerange{\pageref{firstpage}--\pageref{lastpage}}
\maketitle

\begin{abstract}
The study of stellar multiplicity offers important constraints on the structure of the Galaxy as well as stellar and planet formation and evolution. Focusing on the most immediate solar neighbourhood benefits from obtaining both complete and accurate data for reliable statistics.
Our goal is to describe the solar neighbourhood within 10\,pc in terms of multiplicity by evaluating the angular and physical separations, masses, and orbital periods of the systems from the most complete volume-limited sample.
We carried out a comprehensive data compilation from the Washington Double Star catalogue and the literature of all known multiple systems at any separation range, and completed this information with a common proper motion and parallax search with \textit{Gaia} DR3 data.
We also used public astrometric and radial-velocity data to compute orbital solutions of seven pairs.
From a sample of 424 stars and brown dwarfs within 10\,pc we identified 215 of them in 92 systems in double (68), triple (19), quadruple (3), and quintuple (2) configurations. 
All except eight pairs have been resolved.
Their orbital periods range over ten orders of magnitude from about one day to millions of years.
We measured precise mass and companion star fractions at different mass intervals. 
The multiplicity fraction smoothly decreases from $41^{+11}_{-10}$\,\% for stars with $M \ge 0.50\,$M$_\odot$ to $9.3^{+7.4}_{-4.3}$\,\% for stars and brown dwarfs with $M \le 0.10\,$M$_\odot$. 
\end{abstract}




\section{Introduction}
\label{sec:introduction}

The study of stellar multiplicity has interested astronomers for millennia. From Ptolemy in his Almagest in the 2nd century \citep{peters1915,grasshoff90,argyle19}, through Hodierna in the first catalogue of binary stars in 1654 \citep{gonzalezpayo24b}, to modern astronomers who have compiled the most recent and comprehensive lists of stars and brown dwarfs in the solar neighbourhood at $d <$ 10\,pc \citep{henry06,reyle21}, 20\,pc \citep{kirkpatrick24}, and 100\,pc \citep{gaiacollaboration21b}.
Understanding stars and substellar objects in the solar neighbourhood provides insights into galactic structure, stellar formation and evolution, planetary system dynamics, and, as developed below, even the search for habitable worlds. 

The study of resolved stellar multiplicity has evolved with observing techniques (naked eye, position micrometers, seeing-limited imaging from the ground with photographic plates and digital detectors, eyepieces, speckle and multi-aperture interferometry, lucky imaging, adaptive optics, high-resolution imaging from space), as well as with the explored angular separations, at both the closest and widest separation ends \citep{batten73,caballero09,duchene13,tokovinin14b,tokovinin18,dhital15}.
The interest of astronomers in stellar multiplicity has also evolved, and it seems to have reached a plateau just when \textit{Gaia} is releasing the greatest and most accurate astro-photometric datasets in history \citep{gaiacollaboration18, gaiacollaboration23b}. 

There is, however, a renewed interest in stellar multiplicity studies, especially in the solar neighbourhood. 
This fact comes from current searches for Earth-mass exoplanets around nearby M dwarfs and future searches for biosignatures on such exoplanets, but around solar-like stars.
On the one hand, close binaries, used for instance for parameter determination, are in many cases discarded from extreme-precision radial-velocity searches \citep{roell12,cortescontreras17b,fouque18,gonzalezpayo24,cifuentes25}. On the other hand, wide binaries help to test evolutionary models and are of help in the determination of M-dwarf planet-host parameters that are challenging to determine in single, cool stars \citep{montes18,duquearribas24}.
Examples of those parameters are stellar masses (e.g. dynamical masses in a spectro-astrometric binary -- \citealt{gonzalezalvarez20} for GJ~338) and element abundances that are critical for the formation and composition of rocky planets (e.g. Mg and Si in multiple systems with FGK-type primaries and M-type secondaries -- \citealt{tabernero24}).

On the other hand, both NASA and ESA are starting to design their flagship missions for the characterisation of habitable planets, which are set to launch in the 2040--2050 time frame.
NASA's Habitable Worlds Observatory\footnote{\url{https://science.nasa.gov/astrophysics/programs/habitable-worlds-observatory/}} (HWO -- \citealt{clery23,dressing24}) will likely be in orbit first, and will explore exoplanet atmospheres in the ultraviolet, visible, and near infrared with a $\sim$6-m segmented primary mirror.
On the other side of the Atlantic, one of the final recommendations from the ESA's Voyage 2050 Senior Committee was to develop a mission specifically focusing on the `Characterisation of Temperate Exoplanets' science theme\footnote{\url{https://www.cosmos.esa.int/web/voyage-2050}}.
The Large Interferometer For Exoplanets\footnote{\url{https://life-space-mission.com/}} (LIFE -- \citealt{quanz22}), which adequately fits this theme, might be in orbit just afterwards, and complement HWO with interferometric observations in the mid-infrared\footnote{China is also preparing its own exoplanet-hunting space observatory, namely Tianlin (\begin{CJK*}{UTF8}{gbsn}天邻 \end{CJK*}, \citealt{wang23}).} \citep{alei24}.
In order to optimise exoplanet yield simulations and inform the spacecraft optomechanical designs with accurate scientific requirements, both HWO and LIFE need preliminary target lists \citep{mamajek24,tuchow24,harada24,menti24,hartman26}.
Both mission teams emphasise the study of stellar multiplicity, as angular separation and magnitude difference between components are key parameters for current planet yield calculations and the next generation of atmospheric retrieval simulations \citep{konrad22,carriongonzalez23,morgan24}.
Similarly to radial-velocity searches, close binaries will be discarded from future HWO and LIFE programs, while wide multiples will help in determining precise parameters of planet host stars. (the terms `close' and `wide' pairs are often defined using fixed separation thresholds that vary with the study context; nevertheless, this classification should be considered dynamic rather than static, as it depends on the detection capabilities and spatial resolution of the observational facilities employed; \citealt{cifuentes25}). 
In any case, studying Earth-like exoplanets is not the only reason for the renewed interest of astronomers in stellar multiplicity. A few recent examples of astrophysical themes that require stellar multiplicity as a key input are the determination of the local (initial) mass function \citep{kirkpatrick24}, the formation and evolution of ultra-cool dwarfs \citep{baig24}, and the characterisation of white dwarfs \citep{obrien24}.

This is the last item of a trio of papers on a thorough investigation of the widest multiple systems with angular separations $\rho \ge$ 1000\,arcsec \citep{gonzalezpayo23},
the multiplicity of stars with planets at $d <$ 100\,pc \citep{gonzalezpayo24}, 
and of all multiple systems at $d <$ 10\,pc (this work).
The three of them share an identical methodology and the extensive use of data from the \textit{Gaia} third data release (\textit{Gaia} DR3 -- \citealt{gaiacollaboration23b}) and the Washington Double Star catalogue (WDS -- \citealt{mason01}), along with a comprehensive analysis and data compilation from the literature.
The current work should pave the way for new multiplicity surveys \citep{littlefield24,leblanc24,henry24}, 
updates of stellar and substellar catalogues up to 10--20\,pc \citep{reyle21,kirkpatrick24}, 
and ongoing work on stellar multiplicity for HWO and LIFE target lists (C.\,K.~Harada, E.\,E.~Mamajek, F.~Menti, Z.~Hartman, priv. comm.).
A preliminary version of this manuscript was published as part of the PhD thesis of \citet{gonzalezpayo25} and, therefore, benefitted from the comments from a number of individuals listed in the Acknowledgements.

\section{Sample}
\label{sec:sample_10pc}

The list of 541 stars, brown dwarfs, and exoplanets in 336 systems within 10\,pc from the Sun compiled by \citet{reyle22} represents the most complete volume-limited sample based on current knowledge.
This list is the first update to the `The 10~parsec sample in the \textit{Gaia} era' \citep{reyle21}.
The completeness of their compilation has been confirmed later by several authors \citep[e.g.][]{golovin23,kirkpatrick24}, as well as by cross-match with contemporaneous catalogues \citep{hirsch21}.
The latest version of the list of \citet{reyle22} can be retrieved from the VizieR catalogue access system \citep{ochsenbein00} and its dedicated website\footnote{\url{https://gruze.org/10pc/}}.
We only had to delete three entries, corresponding to the L2 and L4 components of the binary 2MASS J06174191+1945135~AB (CWISE J061741.79+194512.8~AB), which have a new spectrophotometric distance estimate of 28.2$\pm$5.7\,pc, far beyond the 10\,pc limit \citep{humphreys23}, and Gaia DR3 6305165514134625024, which is a background object with wrong \textit{Gaia} parallax \citep{golovin23,kirkpatrick24}.
In the compilation by \citet{reyle22}
there are also a few targets with distances slightly farther than $d =$ 10\,pc, but consistent with that value within uncertainties, which we kept in our analysis.

Table~\ref{tab:sample_10pc} displays the default name in the Simbad database, Gliese-Jahreiss identifier when available, J2000 equatorial coordinates, and proper motions of 424 stars and brown dwarfs at less than 10\,pc. 
The \textit{Gaia} early third data release (EDR3; \citealt{gaiacollaboration21a}) was the main provider of astrometric data (identical to that from DR3). 
When EDR3 data are missing, \citet{reyle21,reyle22} compiled them from other sources (e.g. \citealt{vanleeuwen07} for bright stars or close binaries without \textit{Gaia} astrometric solution, but also \citealt{dupuy12} or \citealt{kirkpatrick19,kirkpatrick21} for faint low-mass stars and brown dwarfs).

There are fewer objects catalogued by Simbad than by \citet{reyle22}, which explains the difference in the number of objects in Table~\ref{tab:sample_10pc}.
All of the apparently missing objects are in close multiple systems.
We added square brackets in the names for clarity and easy cross-match with Simbad.
As an example, \citet{reyle22} tabulated $\chi$~Dra~A and $\chi$~Dra~B separately, we show $\chi$~Dra~[AB] in a single row in Table~\ref{tab:sample_10pc}, and Simbad only catalogues $\chi$~Dra.

\section{Analysis and results}
\label{sec:analysis}

\subsection{Common proper motion and parallax pairs}
\label{sec:common}

As mentioned in Sect.~\ref{sec:introduction}, we followed the same methodology as \citet{gonzalezpayo23,gonzalezpayo24}.
First we looked for common proper motion and common parallax \textit{Gaia} DR3 companions to the 424 stars and brown dwarfs in Table~\ref{tab:sample_10pc}.
We imposed maximum relative proper motion modulus and parallax differences of 15\,\% and maximum proper motion angle difference of 15\,deg.
These conservatively large criteria limits were empirically justified by \citet{gonzalezpayo23,gonzalezpayo24} to account for high proper motion projection effects in nearby, very wide multiple systems \citep{wertheimer06} and proper motion and parallax anomalies in very close systems \citep{kervella19,brandt21b}.
We set a maximum projected physical separation of our search of 1\,pc ($s \sim 2 \cdot 10^5$\,au). The reason behind this limit is that the expected number of gravitationally bound systems with a wider separation is very low, as the strength of the galactic tidal field is comparable to their gravitational attraction \citep{binney08}.
The value of 1\,pc is identical to similar searches in the literature \citep[e.g.][]{el-badry21} and slightly shorter than the Jacobi radius of 1.35\,pc for a binary with a total mass of 1\,M$_\odot$.
Binaries more separated than the Jacobi radius are disrupted by gravitational perturbations from other stars and molecular clouds \citep{weinberg87,jiang10}.
We should note that a universal Jacobi radius is a mere approximation of the galactic tidal force. 
For a given total mass, the limiting separation depends also on eccentricity, as well as on the alignment of the orbital momentum axis with the axis of galactic rotation \citep{makarov12}.
There are more sophisticated common proper motion and common parallax searches that include, for example, parallax uncertainties and the assumption of proper motions consistent with a Keplerian orbit \citep[e.g. again][]{el-badry21}.
However, although simpler, our method seems to be superior for very nearby, very wide systems with projection effects, such as $\alpha$~Centauri~AB and Proxima, and for hierarchical triple and quadruple systems with underestimated \textit{Gaia} parallax uncertainties of components that actually are tight binaries \citep{gonzalezpayo23,gonzalezpayo24}.

Next, we complemented our \textit{Gaia} DR3 search with WDS, discarding from the analysis all background and spurious sources (usually with the `U' flag).
All the companions found in our \textit{Gaia} search are already tabulated by WDS.
However, WDS tabulates a number of systems not found in the \textit{Gaia} search, due to either very close separation between components (e.g. G~158--50~A,B, WDS~00155--1608, HEI~299: $\rho \approx$ 0.3037\,arcsec as measured by \citealt{ment23}), or faintness of the companion, with a $G$ magnitude beyond the \textit{Gaia} completeness (e.g. HD~42581~A,Ba,Bb, WDS~06106--2152, NAJ~1: the companion is the famous `GJ~229\,B' T-type brown dwarf discovered by \citealt{nakajima95}).
We discarded only one WDS system, namely Wolf~358 (J10509+0648, RAO~256), which is described in detail in Appendix~\ref{sec:remarkable_systems_10pc}.

We enhanced our \textit{Gaia} and WDS searches with a detailed review of the literature using the Simbad database \citep{wenger00}.
We identified multiple systems that were absent in our \textit{Gaia} search or not tabulated by WDS.
Some of them are unresolved close pairs, including spectroscopic binaries, or have recently-resolved ultra-cool companions (e.g. GJ~229\,Ba,Bb, \citealt{xuan24}).
Most of them were already identified by \citet{reyle22} or \citet{kirkpatrick24}
or listed by the Multiplicity Star Catalog\footnote{\url{https://www.ctio.noirlab.edu/~atokovin/stars/}} (MSC; \citealt{tokovinin18}), which tabulates hierarchical multiple stellar systems with three or more component candidates.
In this literature analysis, we discarded the M3\,V+T9 wide pair L~34--26~AB \citep{zhang21}, whose secondary at $d = 10.22\pm0.70$\,pc is the only component listed in Table~\ref{tab:sample_10pc}; the M-dwarf primary is unambiguously beyond our 10\,pc limit.

\subsection{\textit{Gaia} multiplicity indicators}
\label{sec:gaia}

We used \textit{Gaia} data to supplement the information of our sample of multiple stars.
First, we searched the \textit{Gaia} DR3 Non-Single Stars (NSS) catalogue \citep{gaiacollaboration22b,halbwachs23}.
Six objects in our sample are listed in the \texttt{nss\_two\_body\_orbit} table. 
According to the documented scheme \citep{pourbaix22}, four of them (BD--15~6290, Wolf~227, L~88--59, and 41~Ara\,B) are classified under the `OrbitalTargetedSearch' solution type (orbital model for a priori known systems), one object (G~184--19\,A) is flagged as `SB2' (spectroscopic binary), and one (FK~Aqr\,A) as `OrbitalTargetedSearch/SB2C' (spectroscopic binary with circular orbit).
No counterparts were found in the other NSS tables (\texttt{nss\_acceleration\_astro}, \texttt{nss\_vim\_fl}, or \texttt{nss\_non\_linear\_spectro}).

For Wolf~227 and 41~Ara\,B, NSS does not provide extra information to what is already presented in Sect.~\ref{sec:results_10pc}.
For G~184--19\,A and FK~Aqr\,A, NSS tabulates exactly half and twice the more precise (and accurate) values of orbital periods determined by \citet{delfosse99} and \citet{tsvetkova24}, respectively.
BD--15~6290 is the well-known planet-host M-dwarf star GJ~876 \citep{marcy98,delfosse98}; the period reported by NSS appears to be a non-linear combination of those of the Jovian planets GJ~876\,b ($P_{\rm b}$ = 61.1035\,d) and GJ~876\,c ($P_{\rm c}$ = 30.1039\,d).
Finally, the single white dwarf L~88--59 is known to display spurious multiplicity signatures \citep{rogers24}.

Next, we applied the close-multiplicity criteria to identify unresolved sources based on \textit{Gaia} DR3 statistical indicators summarised in table~3 of \citet{cifuentes25}.
These criteria include the renormalised unit weight error ({\tt RUWE}), which quantifies the goodness of fit between the observed astrometric data and a single-star model \citep{arenou18,lindegren18a}, {\tt ipd\_gof\_harmonic\_amplitude} which was designed to flag spurious solutions in resolved doubles \citep{fabricius21}, and several measures of radial-velocity variability among all the \textit{Gaia} measurement epochs \citep{katz23}.
Although that was its intended purpose, {\tt ipd\_gof\_harmonic\_amplitude} is not an indicator of unresolved binarity in \textit{Gaia} DR3. Due to a processing error, it reflects the degree of photometric variability instead\footnote{\url{https://www.cosmos.esa.int/web/gaia/dr3-known-issues}}.

There are 16 single stars and resolved binary components that satisfy at least one of the multiplicity criteria of \citet{cifuentes25}.
Of them, fifteen are very bright stars with high {\tt RUWE} (e.g. $\tau$~Cet, $\epsilon$~Eri, $\pi^{03}$~Ori), M dwarfs with astrometric signals produced by giant planets (e.g. GJ~876), white dwarfs with spurious radial-velocity or astrometric perturbations (e.g. EGGR~290, L~88--59), or well-investigated stars with wrong duplicated-source labels (e.g. GJ~486).
The only remaining star with a reliable \textit{Gaia} multiplicity indicator is Ross~619, which has a {\tt RUWE} value of 2.097.
This moderate \texttt{RUWE} excess is suggestive of unresolved binarity or otherwise anomalous behaviour of the photocentre.
Nevertheless, Ross 619 is a single star based on high resolution imaging (\citealt{oppenheimer01,hinz02,janson14}; Z.~Hartman, priv. comm.), and spectrography (\citealt{bonfils13}; \citealt{ribas23}; \citealt{mignon24}).
A \texttt{RUWE} value larger than 1.4 \citep{arenou18,lindegren18a} is neither a necessary nor a sufficient condition for unresolved binarity \citep[e.g.][]{castroginard24,cifuentes25}.

We also cross-matched our sample with the `Astrometrically identified nearby binary stars' catalogue of \cite{penoyre22}, resulting in seven matches. 
None of these objects were resolved by \textit{Gaia}, due to the faint companion star's magnitude ($\chi^{01}$~Ori), close separations to their primaries (Wolf~227, G~203--47, and L~768--119), or a combination of both (BD--18~ 359, WT~460, HD~16169; all with $\rho \approx$ 0.4--0.5\,arcsec and $\Delta G \gtrsim$ 4\,mag). 
The astrometric solutions for these objects show consistently low \texttt{RUWE} and local unit weight error (\texttt{LUWE}) values, with only two exceptions, namely BD--18~359 and G~203--47, where notably higher values are observed (\texttt{RUWE} = 15.1 and 23.2, respectively). 
However, in both cases, the results still point to well-known close binarity. 
As a result, both \texttt{RUWE} and \texttt{LUWE} lead to the same conclusion about the binarity of each of our targets.
This is in line with the expected \texttt{RUWE}--\texttt{LUWE} differences at large heliocentric distances \citep{penoyre22} and previous applications of \texttt{LUWE} in the solar neighbourhood \citep{baig24}.
All in all, our analysis of \textit{Gaia} multiplicity indicators did not provide any new close binary, but confirmed some previous classifications.

\subsection{Single stars and brown dwarfs, and completeness}
\label{sec:single}

From our common proper motion and parallax survey and literature compilation, and \textit{Gaia} multiplicity indicators analysis, we identified 215 stars and brown dwarfs in 123 pairs, which are organised into 92 unique systems at $d <$ 10\,pc (Tables~\ref{tab:systems_10pc} and \ref{tab:pairs_10pc}). 
The difference between the number of pairs (123) and systems (92) is due to the hierarchical architecture of the sample: while a binary system contributes a single pair, the 19 triple, 3 quadruple, and 2 quintuple systems identified contribute multiple pairs to the physical distributions, even though each counts as only one system for multiplicity statistics

The remaining 259 stars and brown dwarfs in Table~\ref{tab:sample_10pc}, not displayed in Tables~\ref{tab:systems_10pc} and \ref{tab:pairs_10pc}, are considered to be single with current data, including 2MASS J09393548--2448279, which was originally presented as `the coldest and least luminous brown dwarf binary known' \citep{burgasser08}.
Again, the difference in the total number of objects is due to close multiple systems not resolved by \citet{reyle22}.

All single stars have a \textit{Gaia} counterpart except for two A-type stars due to extreme brightness (Altair, Vega) and 56 ultracool dwarfs due to extreme faintness (1 L, 30 T, 25 Y).
The two early-type stars have been the subject of numerous multiplicity studies (e.g. \citealt{fuhrmann15,bouchaud20}), while some of the ultracool dwarfs still require further multiplicity analysis (e.g. \citealt{bardalezgagliuffi14}).
However, considering their current estimated multiplicity fractions, between 12\% and 9\% for L and T respectively \citep{burgasser03c,reid08b,dalponte20}, no more than two T new binaries would be expected. 

Of the 201 remaining single stars and brown dwarfs with \textit{Gaia} counterpart, 26 are FGK stars, 143 are M dwarfs, 13 are white dwarfs, 13 are L dwarfs, and 6 are T dwarfs.
A large fraction of the single FGKM stars have been extensively studied with extreme precision spectrographs \citep[e.g.][]{mayor11,howard16,wittenmyer16,ribas23,mignon24,harada25} and with high-resolution, high-contrast imagers \citep[e.g.][]{oppenheimer01, chauvin10, vigan21, gauza21}, besides with \textit{Gaia}.
In particular, 97 of the 143 M dwarfs have been monitored by radial-velocity surveys, mainly with CARMENES \citep{quirrenbach14} in the northen hemisphere and the celestial equator, but also with HARPS \citep{bonfils13}, SPIRou \citep{moutou24}, and even UVES \citep{zechmeister09} in the rest of the sky.
The confirmation of the very few hypothetical single stars and brown dwarfs that have not yet been the subject of detailed multiplicity studies, 
such as the M5.5\,V star PM~J18057--1422 at $d \approx$ 9.5\,pc \citep{alonsofloriano15a}, will be the topic of forthcoming work (\'E. Artigau, priv. comm.).
We delayed an exhaustive multiplicity sample completeness analysis to after the completion of on-going and upcoming high-resolution imaging and spectroscopic surveys at large facilities, such as with the 10.4\,m Gran Telescopio Canarias Adaptics Optics system \citep[GTCAO,][]{bejar20}, now being commissioned. 
Hereafter, we assumed that past surveys have identified all multiple systems within 10\,pc, that we were able to identify and discard all spurious multiple systems, and, therefore, that all stars and brown dwarfs classified by us as `single' do not have any companion down to the deuterium-burning mass limit at about 0.013\,M$_\odot$, widely used as the brown dwarf-planet boundary.

\subsection{Binary and multiple systems}
\label{sec:results_10pc}

Tables~\ref{tab:systems_10pc} and~\ref{tab:pairs_10pc} summarise our analysis and data compilation. 
In particular, Table~\ref{tab:systems_10pc} lists the individual components in all multiple systems at $d <$ 10\,pc (e.g. two entries for a double system --A, B--, three entries for a triple system --A, B, C--, etc.), while Table~\ref{tab:pairs_10pc} lists the pairs (e.g. one entry for a double system --AB--, two entries for a triple system --A-B and A-C--, etc.).
The first column, the WDS identifier, is common to both tables.
Of the 92 multiple systems at $d <$ 10\,pc, only five do not have a WDS identifier, namely Wolf~227, QY~Aur, L~768--119, G~203--47, and 2M1750--00 (2MASS J17502484--0016151).
We write their names in parentheses in Tables~\ref{tab:systems_10pc} and~\ref{tab:pairs_10pc}.

For every component in each system, Table~\ref{tab:systems_10pc} lists:
\begin{itemize}
\item WDS discoverer code,
\item default Simbad name of star or brown dwarf (in square brackets, component letter added by us),
\item alternative name or names (only for stars with bright, variable, historical, or widely accepted denominations),
\item GJ identifier \citep{gliese79,gliese91},
\item component letter used by WDS or MCS (in square brackets, component not identified by any of these catalogues),
\item ICRS coordinates (provided only for the brightest component when unresolved by \textit{Gaia}),
\item heliocentric distance from \citet{reyle22}, in turn compiled from \textit{Gaia} (E)DR3 and a number of heterogeneous sources,
\item spectral type (the colon stands for spectral types estimated by us from photometry, 14\,\%),
\item \textit{Gaia} $G$ magnitude (the colon stands for $G$-band magnitudes estimated by us from previously-published magnitude differences in other optical and near-infrared bands, 26\,\%),
\item mass $M$ in solar masses (masses compiled from the literature or estimated by us from absolute magnitudes and the relations or models of \citealt{pecaut13} for two K dwarfs, \citealt{cifuentes20} for four M dwarfs, and \citealt{chabrier00} for three L dwarfs, or determined by us from astrometric and radial-velocity orbital fits as described in Sect.~\ref{sec:orbfit}), 
\item bibliographic reference of the mass (18\,\% from \citealt{kirkpatrick24}, 57\,\% from \citealt{cifuentes25}, 6\,\% from `This work', and 19\,\% from other references -- \citealt{burgasser00,gelino11,liu12,vigan12,pope13,wright13,zhang21,bedin24,xuan24}), and
\item reported planet candidates (updated results of \citealt{gonzalezpayo24} with the Encyclop{\ae}dia of exoplanetary systems\footnote{\url{https://exoplanet.eu/}} --\citealt{schneider11}-- and NASA Exoplanet Archive\footnote{\url{https://exoplanetarchive.ipac.caltech.edu/}} --\citealt{akeson13}--; 
in square brackets, unconfirmed planet candidates).
\end{itemize}
Besides, Table~\ref{tab:systems_10pc} also lists for each system its reduced binding energy modulus $|U_g^*|$ in $10^{33}$\,J, usually measured between the widest component and the rest of the components in the system \citep{caballero07a}, and a diagram sketch (the symbol and colour legend is provided in the table footnote).

For every pair in each system, Table~\ref{tab:pairs_10pc} lists:
\begin{itemize}
\item WDS discoverer code,
\item letter of `primary' and `secondary' component used by WDS or MCS (in square brackets, component not identified by any of these catalogues),
\item detection method 
(\textit{Gaia}: \textit{Gaia} DR3; 
DI: Direct seeing-limited imaging --not in \textit{Gaia} DR3--; 
HRI: High-resolution imaging --\textit{Hubble}, adaptive optics, speckle, interferometry, etc.--; 
SB1/SB2: Single-/double-lined spectroscopic binarity),
\item position angle $\theta$ (measured from North to East; `--' when $\rho <$ 1\,arcsec), 
\item angular separation from primary to companion $\rho$ in arcsec,
\item projected physical separation in au ($s \approx \rho \cdot d$; exact formula in eq.~6 of \citealt{gonzalezpayo23}),
\item reference for $\theta$ and $\rho$ (`This work' for 21 of the 114 resolved pairs; new $\theta$ and $\rho$ values were measured by us from \textit{Gaia} DR3 astrometric data at epoch J2016.0), 
\item angular semi-major axis of the orbit $\alpha$ in arcsec (in italics, derived from $a$ by us),
\item spatial semi-major axis of the orbit $a$ in au (in italics, derived from $\alpha$ by us),
\item orbital period $P$,
\item unit of orbital period (d: days, a: years, ka: thousand years, Ma: million years),
\item orbital grade from the Sixth Catalog of Orbits of Visual Binary Stars\footnote{\url{https://crf.usno.navy.mil/wdsorb6}} (ORB6) of the WDS, which estimates the quality of period measurement (`--' when not provided; `$<$' when improved with respect to ORB6), and
\item reference for $\alpha$, $a$, or $P$.
\end{itemize}

Pairs in Table~\ref{tab:pairs_10pc} can be split into two classes: 
those with orbital fits to astrometric, spectroscopic, or spectro-astrometric data (with $\alpha$, $a$, or $P$ -- we do not tabulate $\theta$, $\rho$, $s$ for them), and those without orbital fits (but with $\theta$, $\rho$, $s$).
For the latter, we estimated $P$ from $M_i$ and $M_j$ in Table~\ref{tab:systems_10pc}, $s_{ij}$ in Table~\ref{tab:pairs_10pc}, and Kepler's third law, $P_{ij}^2 = (M_i + M_j)^{-1} ~ a_{ij}^3$ in Solar System units.
We adopted a statistical conversion between projected separation and semimajor axis, $a \approx 1.26\,s$, following previous works \citep{abtlevy76,
fischer92,torres99a,close03,burgasser07a,radigan09,caballero10,faherty10,dupuyliu11}.
We warn, however, that new simulations of the projection effect involving specific models of eccentricity reveal a broad convolution kernel and, therefore, quite uncertain actual value of $a$ without an accurate 3D orbit.
For example, \cite{makarov25} concluded that the expectation (or the sample mean $a$) is not very sensitive to distribution of $e$, with scaling transformations between $a \approx 1.06 s$ for mostly small and moderate $e$ (close binaries) and $a \approx 0.96 s$ for $e$ piling up at 1 (wide binaries).
This correcting factor is needed to homogeneously compare `true' $P$ that are determined from orbital fits and `approximated' $P$ that are estimated from stellar masses and physical separations without a priori knowledge of orbital parameters ($e$, $i$, $\omega$, $\Omega$).
However, we computed the $|U_g^*|$ values in Table~\ref{tab:systems_10pc} with $s$, which, therefore, can be compared among themselves and with many other published work \citep{close03,burgasser07a,caballero10,gonzalezpayo21,gonzalezpayo23,gonzalezpayo24,cifuentes25}.

As mentioned before, not all nearby multiple systems have been resolved. There are eight very close unresolved short-period ($P$ = 1.8--62.6\,d) spectroscopic binaries at $d <$ 10\,pc.
Six of them are in hierarchical systems, including three triples, one quadruple, and the two quintuples. 
Four of the hierarchical systems and one of the `single' spectroscopic binaries that has a poorly characterised white-dwarf component candidate are described, with the corresponding references, in Appendix~\ref{sec:remarkable_systems_10pc}.

\subsection{New orbital solutions}
\label{sec:orbfit}

\begin{table*}
\centering
\caption{New orbital solutions of seven wide pairs in multiple systems at $d < 10$\,pc.}
\scalebox{0.9}[1]{
\begin{tabular}{lccccccc}
\hline
\noalign{\smallskip}

WDS & 00321+6715 & 03019--1633 &  05404+2448 & 05544+2017 & 14545+1606 & 19539+4425 & 20112--3606 \\

\noalign{\smallskip}
\hline
\noalign{\smallskip}

Discovery code &
VYS 2 &
RST 2292 &
WNO 45 &
KNG 1 &
FRT 1 &
GIC 159 &
HJ 5173 \\
\noalign{\smallskip}
\noalign{\smallskip}
Primary &
BD+66 34Aa,Ab &
BD--17 588A &
G 100--28[A] &
$\chi^{01}$ Ori [A] &
BD+16 2708[A] &
G 208--44[AB] &
HD 191408[A] \\
\noalign{\smallskip}
\noalign{\smallskip}
Secondary &
BD+66 34B &
BD--17 588[BC] &
G 100--28[B] &
$\chi^{01}$ Ori [B] &
BD+16 2708B[a,b] &
G 208--45 &
HD 191408B \\
\noalign{\smallskip}
\noalign{\smallskip}
$N_{\rm obs}^{\rm ast}$ &
107 & 13 & 9 & 3 & 13 & 22 & 20 \\
\noalign{\smallskip}
\noalign{\smallskip}
$N_{\rm obs}^{\rm RV}$ &
 -- & 232 & -- & 246 & 48 & -- & 436 \\
\noalign{\smallskip}
\noalign{\smallskip}
Method &
 {\tt H\&M} & \texttt{orvara} & {\tt H\&M} & \texttt{orvara} & \texttt{orvara} & {\tt H\&M} & \texttt{orvara} \\
\noalign{\smallskip}
\noalign{\smallskip}
$P$ [a] &
$311^{+26}_{-21}$ &
${216}_{-11}^{+14}$ &
$35^{+64}_{-20}$ &
$14.098^{+0.010}_{-0.010}$ &
$440^{+42}_{-34}$ &
$342^{+61}_{-41}$ &
$591^{+118}_{-84}$ \\
\noalign{\smallskip}
\noalign{\smallskip}
$T_0$ [a] &
$1990.4^{+1.9}_{-0.9}$ &
${2116.5}_{-5.0}^{+6.2}$ &
$2010.6^{+5.8}_{-4.0}$ &
$2013.853^{+0.024}_{-0.024}$ &
$2104^{+60}_{-55}$ &
$1971^{+13}_{-17}$ &
$2213^{+111}_{-63}$ \\
\noalign{\smallskip}
\noalign{\smallskip}
$\alpha$ [arcsec] &
$4.138^{+0.0056}_{-0.0058}$ &
${4.692}_{-0.193}^{+0.227}$ &
$0.73^{+0.45}_{-0.05}$ &
$0.719^{+0.006}_{-0.006}$ &
$5.02^{+0.23}_{-0.20}$ &
$7.89^{+3.28}_{-0.15}$ &
$11.6^{+1.6}_{-1.3}$ \\
\noalign{\smallskip}
\noalign{\smallskip}
$e$ &
0 (fixed) &
$0.561^{+0.068}_{-0.072}$ &
$0.22^{+0.34}_{-0.12}$ &
$0.454^{+0.003}_{-0.002}$ &
$0.055^{+0.032}_{-0.030}$ &
0 (fixed) &
$0.75^{+0.10}_{-0.10}$ \\
\noalign{\smallskip}
\noalign{\smallskip}
$i$ [deg] &
$54.2^{+3.9}_{-4.5}$ &
$87.27^{+0.29}_{-0.34}$ &
$50.1^{+10.9}_{-18.7}$ &
$92.69^{+0.39}_{-0.39}$ &
$11.0^{+6.6}_{-4.5}$ &
$121.2^{+5.5}_{-14.1}$ &
$86.79^{+0.60}_{-0.86}$ \\
\noalign{\smallskip}
\noalign{\smallskip}
$\omega$ [deg] &
0 (fixed) &
${350.0}_{-1.3}^{+1.3}$ &
$186^{+98}_{-67}$ &
$110.74^{+0.42}_{-0.47}$ &
$129^{+67}_{-64}$ &
0 (fixed) &
$66.2^{+13.0}_{-6.7}$ \\
\noalign{\smallskip}
\noalign{\smallskip}
$\Omega$ [deg] &
$164.1^{+1.3}_{-0.58}$ &
${135.358}_{-0.069}^{+0.078}$ &
$127^{+15}_{-21}$ &
$306.51^{+0.52}_{-0.52}$ &
$30^{+42}_{-22}$ &
$98.9^{+9.4}_{-7.0}$ &
$294.4^{+1.8}_{-2.6}$ \\
\noalign{\smallskip}
\noalign{\smallskip}
$M_{\rm primary}$ [M$_\odot$] &
 -- & $0.286^{+0.018}_{-0.018}$ & -- & $1.075^{+0.033}_{-0.033}$  & $0.493^{+0.046}_{-0.039}$ & -- & $0.750^{+0.030}_{-0.030}$ \\
\noalign{\smallskip}
\noalign{\smallskip}
$M_{\rm secondary}$ [M$_\odot$] &
 -- & $0.431^{+0.017}_{-0.017}$ & -- & $0.157^{+0.003}_{-0.003}$ & $0.150^{+0.004}_{-0.003}$ & -- & $0.207^{+0.046}_{-0.042}$ \\

\noalign{\smallskip}
\hline
\end{tabular}
}
\label{tab:new_orbital_solutions}
\end{table*}

We determined new orbital solutions for seven pairs, listed in Table~\ref{tab:new_orbital_solutions}.
They are resolved by \textit{Gaia} and had estimated orbital periods of less than 1000\,a and very low-grade orbital solutions or no orbital solutions at all.
One of the systems, namely RST~2292 in the system WDS~03019--1633, 
contains a celebrated transiting planet (BD--17~588A\,b, also known as LTT~1445A\,b -- \citealt{winters19,winters22,diamond-lowe23,lavie23,bennett25}).
Likewise, the pair FRT~1 is part of a well-investigated trio of intermediate and late M dwarfs, WDS~14545+1606, one of which is very likely a brown dwarf (namely GJ~659\,Bb). 
In Appendix~\ref{sec:remarkable_systems_10pc} we provide additional details on this remarkable system.

As a first step of our orbital analysis of the seven pairs, we corrected the $\theta$ values from equinox precession \citep{vandekamp1967}.
However, since the proper-motion correction of the position angle for the oldest data is much smaller than the typical uncertainty, we did not correct the $\theta$ values from proper motion.

Four pairs are part of triple systems and have high-grade orbital solutions of the inner pairs, with periods of 2.4--53.6\,a.
In triple systems, WDS tabulates a heterogeneous collection of astrometric measurements of the wide components with respect to the primaries of the inner pairs or, more often, to the photocentroid of the inner pairs.
We homogeneously reduced all astrometric measurements to positions referred to the photocentre of the inner pair.
We determined the photocentre from the barycentre and the centre of fluxes, computed from the mass and flux ratios, $\mathcal{B}$ and $\beta$, defined as:
\begin{equation}
\mathcal{B} = \frac{M_{\rm Ab}}{M_{\rm Aa}+M_{\rm Ab}},
\quad
\beta = \frac{1}{1+10^{-0.4 \Delta m}},
\end{equation}
\noindent where $\Delta m = m_{\rm Ab} - m_{\rm Aa}$ is the magnitude difference between the components.
The inner-pair masses were taken from the data compilation in Sect.~\ref{sec:results_10pc}, and the magnitude differences $\Delta m$ from the WDS dataset. 
Since $\Delta m$ is wavelength dependent, $\beta$ depends on the photometric band; for example, in WDS~00321+6715, $\Delta m$ varies from 0.06\,mag in the optical blue to 0.13\,mag in the $K$ band.

\begin{figure*}
\centering
\hspace{-3.5cm}
\begin{minipage}[t]{0.49\textwidth}
    \vspace{4pt}
    \centering
    \includegraphics[width=0.75\linewidth]{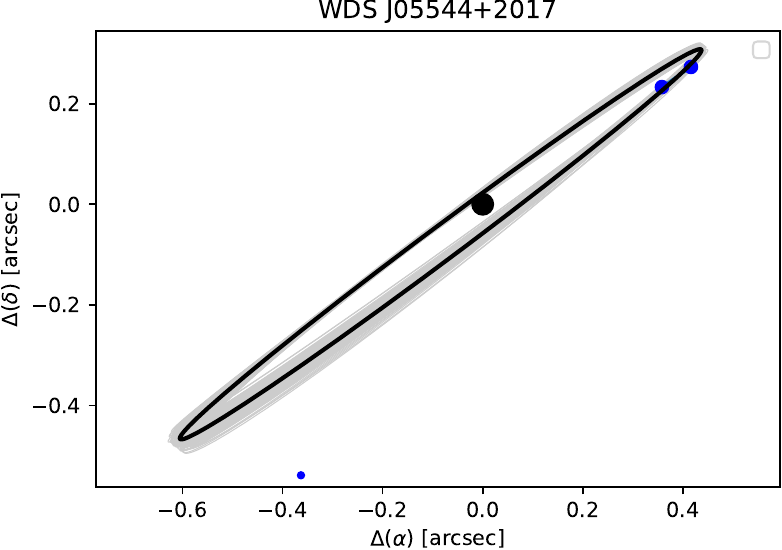}
    \vspace{0.5cm}
    \includegraphics[width=0.73
    \linewidth]{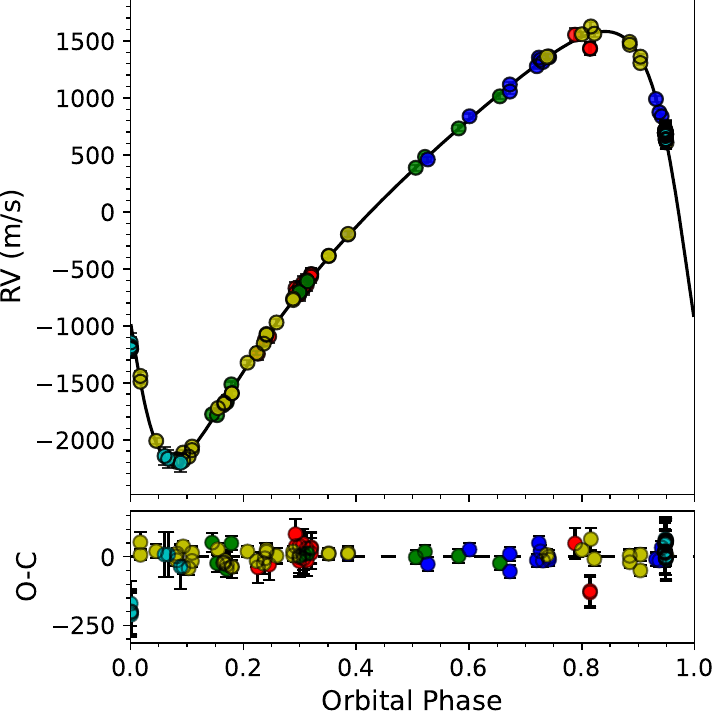}
\end{minipage}
\hspace{-1cm}
\begin{minipage}[t]{0.49\textwidth}
    \vspace{0pt}
    \centering
    \includegraphics[height=0.48\textheight]{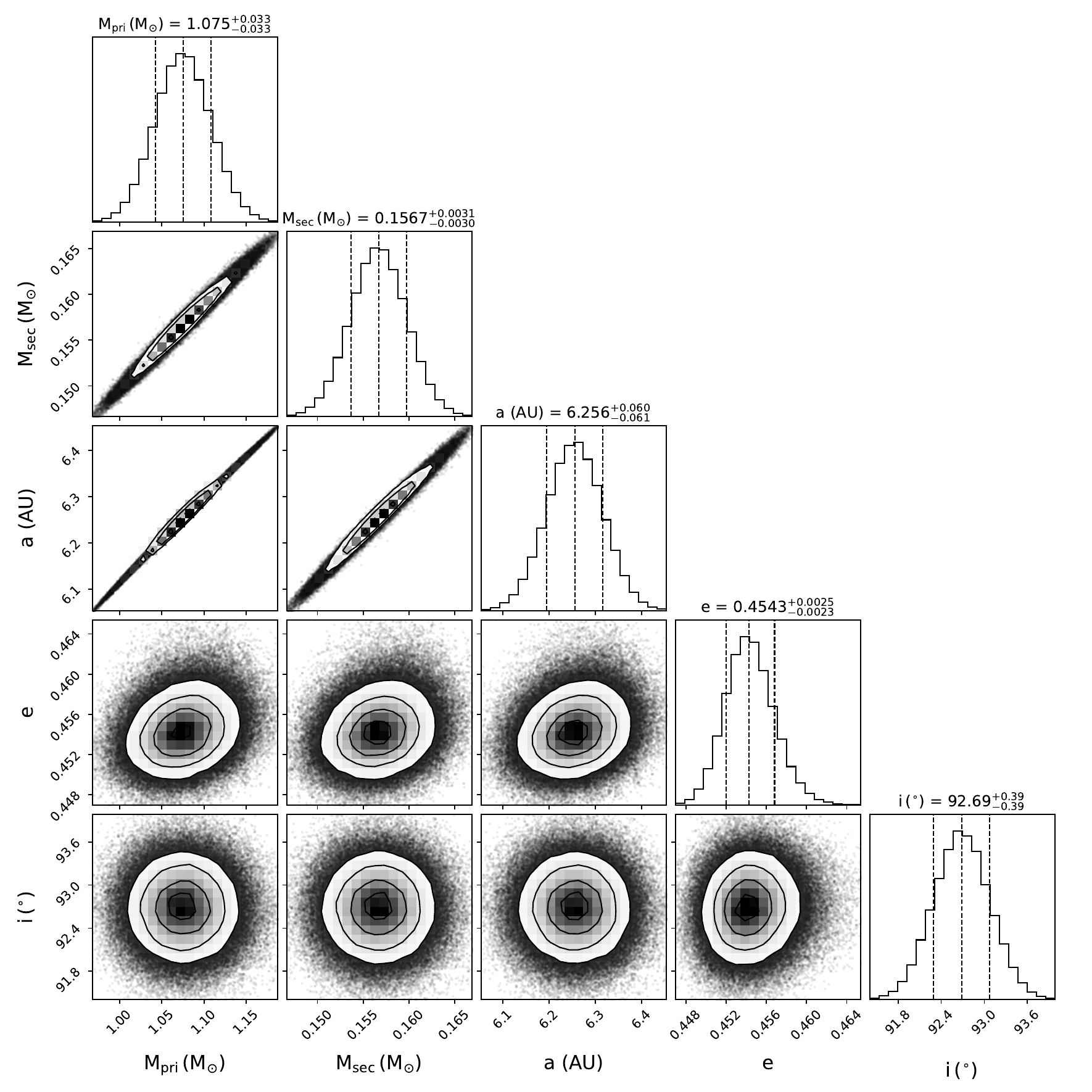}
\end{minipage}
\caption{Orbital fit of $\chi^{01}$\,Ori (KNG~1, WDS J05544+2017) with {\tt orvara}.
\textit{Top left}: astrometric fit (a full legend is in Fig.~\ref{fig:10pc_orbital_solutions}; the outlier data point corresponds to epoch 1982).
\textit{Bottom left}: radial-velocity fit (red: Hamilton-pre, green: Hamilton-post, cyan: Gecko, blue: ELODIE, yellow: SOPHIE).
\textit{Right}: cornerplot with $M_{\rm primary}$, $M_{\rm secondary}$, $a$, $e$, and $i$.} 
\label{fig:chi01_Ori}
\end{figure*}

We used two different methodologies for the orbital fitting depending on data availability and orbital period.
For four pairs, namely RST~2292 in WDS~03019--1633, KNG~1 in WDS~05544+2017, FRT~1 in WDS~14545+1606, and HJ~5173 in WDS~20112--3606, we used the {\tt orvara} open-source Python package \citep{brandt21b} for simultaneous orbital fits with both relative astrometric and radial-velocity measurements, accompanied by the absolute astrometry of the \textit{Hipparcos-Gaia} Catalog of Accelerations \citep{brandt21a} and wide priors of stellar masses as \citet{gonzalezalvarez20}.
In particular, we used:
\begin{itemize}

\item for RST~2292 (Fig.~\ref{fig:chi01_Ori}): 13 astrometric measurements between 1935 and 2020 of the M3.5+M3.0 pair BD--17~588 (LTT~1445), and 232 radial-velocity measurements obtained between 2004 and 2021 with
HARPS\footnote{\url{https://www.eso.org/sci/facilities/lasilla/instruments/harps.html}}, 
HIRES\footnote{\url{https://www2.keck.hawaii.edu/inst/hires/}},  
PFS\footnote{\url{https://users.obs.carnegiescience.edu/crane/pfs/}},
MAROON-X\footnote{\url{https://www.gemini.edu/instrumentation/maroon-x}}, and
ESPRESSO\footnote{\url{https://www.eso.org/sci/facilities/paranal/instruments/espresso.html}}, 

\item for KNG~1: just 3 astrometric measurements between 1982 and 2021 of the G0\,V+M4.5\,V pair $\chi^{01}$~Ori, and 246 radial-velocity measurements obtained between 1982 and 2016 with Gecko\footnote{\url{https://www.cfht.hawaii.edu/Instruments/Spectroscopy/Gecko/}} at the Canada-France-Hawai'i Telescope \citep{irwin92}, Hamilton before and after its upgrade in 1994, Harvard \& Smithsonian's CfA Digital speedometer\footnote{\url{http://astro.vaporia.com/start/cfadigitalspeedometer.html}}, ELODIE\footnote{\url{http://atlas.obs-hp.fr/elodie/intro.html}}, and SOPHIE\footnote{\url{https://ohp.osupytheas.fr/sophie-echelle-spectrograph/}},

\item for FRT 1: 13 astrometric measurements between 1985 and 2021 of the M3.0\,V primary BD+16~2708 [A], and 48 radial-velocity measurements obtained between 2001 and 2013 with HARPS, HIRES, SOPHIE, and ELODIE, and

\item for HJ~5173: 20 relative astrometric measurements between 1896 and 2013 of the K2\,V primary HD~191408, and 436 radial-velocity measurements obtained between 1995 and 2024 with CES\footnote{\url{https://www.eso.org/public/teles-instr/lasilla/coude/ces/}} [both Long Camera and Very Long Camera], HARPS, UCLES\footnote{\url{https://aat.anu.edu.au/science/instruments/decommissioned/ucles/overview}} [University College London Echelle Spectrograph], HIRES, and ESPRESSO.
\end{itemize}

For the purely-astrometric orbital solutions of the pairs VYS~2 in WDS~00321+6715, WNO~45 in WDS~05404+2448, and GIC~159 in WDS~19539+4425, we used the three-dimensional adaptive grid search algorithm developed by \citet{hartkopf89}, and modified afterwards by \citet{mason99}.
We implemented the algorithm in python and called it {\tt H\&M} in honour to Bill Hartkopf and Brian Mason. 
The orbit‐fitting routine generates about $2 \cdot 10^5$ trial orbits by sampling the period ($P$), time of periastron passage ($T_0$), and eccentricity ($e$) over a predefined grid and by solving  the remaining orbital elements via least‑squares at each grid point. 
In two cases (VYS~2 in WDS~00321+6715 and GIC~159 in WDS~19539+4425), the initial best-fit solutions yielded eccentricities compatible within $1\sigma$ with $e = 0$. 
We therefore recomputed the orbital solutions fixing the eccentricity to zero.

We computed the formal uncertainties of the orbital parameters from the set of grid solutions using
    $\Delta \chi^2_i = \chi^2_i - \min(\chi^2)$,
and adopted the condition $\Delta \chi^2_i < 1$ to define the 68.3\% (1$\sigma$) confidence interval. 
The $\chi^2$ statistic was computed from the observed angular separation $\rho$, and position angle $\theta$, transformed into Cartesian coordinates ($x = \rho \cos{\theta}$, $y = \rho \sin{\theta}$).
The quoted error bars correspond to the minimum and maximum values of each parameter within the subset of solutions satisfying $\Delta \chi^2 < 1$. 
Each measurement was represented by a 2D residual vector:
\begin{equation}
\Delta r_i =
\begin{pmatrix}
x_{\rm obs,i} - x_{\rm comp,i} \\
y_{\rm obs,i} - y_{\rm comp,i}
\end{pmatrix},
\end{equation}
and an associated 2$\times$2 covariance matrix:
\begin{equation}
C_i =
\begin{pmatrix}
\sigma_{x,i}^2 & \mathrm{cov}_{xy,i} \\
\mathrm{cov}_{xy,i} & \sigma_{y,i}^2
\end{pmatrix},
\end{equation}
obtained by propagating the observational uncertainties in $\rho$ and $\theta$. 
The total $\chi^2$ was then computed as the sum of Mahalanobis distances \citep{mahalanobis36}:
\begin{equation}
\chi^2 = \sum_i \Delta r_i^\top \, C_i^{-1} \, \Delta r_i.
\end{equation}
\noindent The reduced chi-square was defined as $\chi_{\text{red}}^2 = {\chi^2}/{\nu}$, where the number of degrees of freedom is
$\nu = N_{\rho}+N_{\theta}-7$, with $N_\rho$ and $N_\theta$ being the number of separation and position-angle measurements, respectively, and seven the number of fitted orbital elements.
Values of $\chi^2_{\rm red}$ close to unity indicate a statistically consistent fit and a realistic estimate of the observational uncertainties.

We show the orbital solutions of the seven pairs in Table~\ref{tab:new_orbital_solutions}, which provides WDS system identifier, discoverer code, names of the `primary' and `secondary' stars, number of astrometric and radial-velocity observations, used method ({\tt orvara} or {\tt H\&M}), orbital period ($P$, between about 35\,a and 590\,a), time of periastron passage ($T_0$), angular semimajor axis ($\alpha$), eccentricity ($e$), inclination of the orbital plane ($i$), argument of periapsis ($\omega$), and longitude of the ascending node ($\Omega$) in J2000.0 equinox.
The values of $\Omega$ for the pairs with orbits from relative astrometry only are ambiguous by $\pm$180\,deg (we tabulate $\Omega <$ 180\,deg by convention), which together with the short investigated orbital arcs, makes any study of dynamical stability of triple systems unreliable \citep{hilditch01,mardling01,mardling02}.

However, as a bonus, in addition to determining orbital parameters for the first time, the joint fit with {\tt orvara} allowed us to determine dynamical masses of the components of the investigated systems, identical within uncertainties to the masses estimated from spectro-photometry or determined in the literature.
We report the masses of `primaries' and `secondaries' of systems with {\tt orvara} solutions at the bottom of Table~\ref{tab:new_orbital_solutions}.
In two cases, the `secondaries' are in turn close binaries (BD--17~588[BC] and BD+16~2708B[a,b]).

\section{Discussion}
\label{sec:discussion_10pc}

\subsection{Comparison with previous work}
\label{sec:comparison}

Six pairs in Tables~\ref{tab:systems_10pc} and~\ref{tab:pairs_10pc} are compiled in the Ninth Catalogue of Spectroscopic Binary Orbits (SB9)\footnote{\url{https://sb9.astro.ulb.ac.be/}} of \citet{pourbaix04}.
In all cases, the SB9 periods and those compiled by us from ORB6 or the literature match within uncertainties, except for BD--18~359\,AB.
\citet{nidever02} determined $P$ = 18.7$\pm$6.8\,a from spectroscopic observations, but we kept the most precise value of $P$ = 13.328$\pm$0.037\,a from astrometric observations by \citet{mann19}.

There are 92 multiple systems known at $d <$ 10\,pc.
Of them, 68 are double, 19 triple, 3 quadruple, and 2 quintuple.
We compared our multiple system classification with those of \citet{reyle22} and \citet{kirkpatrick24}, and found a few differences:
WDS~J00363+1821 was single for \citet{reyle22}, but double for \citet{kirkpatrick24} and for us;
WDS~J06106--2152 was double for both \citet{reyle22} and \citet{kirkpatrick24}, but triple for us;
WDS~J06523--0510 was triple for \citet{reyle22}, but double for \citet{kirkpatrick24} and for us.
All of them are described again in Appendix~\ref{sec:remarkable_systems_10pc}.

We went further in our comparison with previous work by cross-matching our tables with the exhaustive MSC.
There are only two systems in our Table~\ref{tab:systems_10pc} that are not listed by MSC, namely WDS~06106--2152 (including the new binary T dwarf GJ~229Ba,Bb) and WDS~22577--2937 (including Fomalhaut and its ultra-wide companions).
While we used MSC for improving the orbital data compilation in Table~\ref{tab:systems_10pc}, we left our previously found parameters when they were already up to date (e.g. \citealt{feng21} instead of \citealt{heintz94} for WDS~02361+0653~AB, HD~16160~AB; \citealt{mason21a} instead of \citealt{izmailov19} for WDS~04153--0739~BC, $o^{02}$~Eri~BC; \citealt{akeson21} instead of \citealt{pourbaix16} for WDS~14396--6050~AB, $\alpha$~Cen~AB).
For the outer systems with very long periods, of tens of thousands of years or longer, for which there are no astrometric periods measured, we also left our own $P$ estimations.
Not surprisingly, our $P$ values are in average about 1.26 longer than those estimated by MSC, which reinforces our projected physical separation and individual masses determinations.
Besides, MSC proposes WDS~14545+1606 to be a quadruple system based on an unconfirmed proper-motion anomaly of the primary (BD+16~2708), but we kept it as a triple (Appendix~\ref{sec:remarkable_systems_10pc}).

Finally, we also checked whether some of our binary systems were also in MSC.
There were eight such multiple (triple, quadruple, quintuple) candidate systems in MSC, but we kept classifying all of them as binaries for a number of reasons (e.g. WDS~04312+5858 and WDS~09144+5241 have been exquisitely investigated in the CARMENES survey by \citealt{gonzalezalvarez20} and \citealt{ribas23}, among many others;
WDS~20298+0941 is a WDS bogus;
the large {\tt RUWE} of the primary of WDS~05445--2227, namely $\gamma$~Lep, is due to its extreme brightness; etc.).
Only remained a quintuple visual system made of a triple in the background including 17~Lyr, and our double in the foreground, which we tabulated as WDS~19074+3230~Ca,Cb.

Further comparison with previous in-depth multiplicity surveys of the solar neighbourhood was out of the scope of this work, since we already included, although indirectly, their input in our meta-analysis \citep[e.g.][]{simons96,schroeder00,oppenheimer01,mccarthy04,carson11,dieterich12,gauza21}.
However, we noted a heterogeneity in the degree of depth and spatial resolution of previous imaging surveys depending on target brightness, spectral type, and heliocentric distance, which may be alleviated in the future by other teams (e.g. \citealt{hartman26}).

\subsection{Multiplicity fraction and companion star fraction}
\label{sec:multiplicity_10pc}

\begin{table}
 \centering
 \caption{MF and CSF as a function of initial stellar mass.}
 \begin{tabular}{l@{\hspace{10mm}}c@{\hspace{10mm}}c}
 \hline 
 \noalign{\smallskip}
 $M_{\rm initial}$ & MF & CSF  \\
 {[M$_\odot$]}  & [\%] &  \\
 \noalign{\smallskip}
 \hline
 \noalign{\smallskip} 
 3.6--0.50 & $41_{-10}^{+11}$ & $0.63_{-0.11}^{+0.10}$ \\
 \noalign{\smallskip}
 \noalign{\smallskip}
 0.50--0.25 & $31.8_{-8.9}^{+10.5}$ & $0.41_{-0.10}^{+0.11}$ \\
 \noalign{\smallskip}
 \noalign{\smallskip}
 0.25--0.10 & $25.3_{-8.0}^{+10.0}$ & $0.310_{-0.088}^{+0.104}$ \\
 \noalign{\smallskip}
 \noalign{\smallskip}
 0.10--0.010 & $9.3_{-4.3}^{+7.4}$ & $0.093_{-0.043}^{+0.074}$ \\
 \noalign{\smallskip}
 \hline
 \end{tabular}
 \label{tab:MF_CSF} 
\end{table}

From the total number of individual stars and multiple systems (binary and higher-order) identified, we determined the multiplicity fraction (MF) and companion star fraction (CSF) following the definitions of \citet{batten73} and \citet{reipurth93}, and recently summarised by \citet{cifuentes25}. 
While the MF represents the fraction of systems in which the primary star belongs to the studied volume, the CSF measures the average number of companions per system. 
We computed MF and CSF strictly per primary mass range and both double and hierarchical systems contribute only one data point to the fractions.
As emphasised by \citet{batten73} and \citet{reipurth93}, without proper primary identification and a clear definition of the counting unit, direct comparisons between these fractions and the overall stellar population may result in misleading interpretations. For this reason, any reference to the local stellar population in the following discussion is intended to be qualitative and illustrative, rather than a strict statistical comparison.

For the whole sample, we measured MF$_{\rm all}$ = $26.2^{+4.8}_{-4.3}$\,\% and CSF$_{\rm all}$ = $0.350^{+0.051}_{-0.048}$, where the uncertainties are 95\,\% Wilson's confidence levels.
However, the original MF and CSF definitions by \citet{batten73} correspond to the mass of the primaries.
When restricted to systems with M-dwarf primaries only, the revised MF$_{\rm M}$ and CSF$_{\rm M}$ values become $29.2^{+6.6}_{-5.8}$\,\% and $0.371^{+0.068}_{-0.064}$, respectively (we did not count systems with white-dwarf components because their stellar progenitors were more massive than M).
Since M dwarfs are by far the most abundant type of stars in the solar neighbourhood \citep{henry06,reyle21,gaiacollaboration21b}, the MF$_{\rm M}$ and CSF$_{\rm M}$ values are in line with other analyses of multiplicity of low-mass stars \citep{henry91,reid97c,janson12,jodar13,cortescontreras17b,winters19,susemiehl22,clark24}.
Including new binary candidates identified in a similar way to this work, \citet{cifuentes25} reported higher MF$_{\rm M}$ and CSF$_{\rm M}$, which supports their hypothesis that their sample was biased by previously unknown unresolved binaries.
These findings further emphasise the need to perform multiplicity statistics on well-defined volume-limited samples and to monitor new \textit{Gaia} DR3 binary candidates with high-resolution imaging and spectroscopic facilities.

We went further in our analysis and determined fractions per mass bin.
We used the data in Table~\ref{tab:systems_10pc} to define four stellar mass intervals, roughly equispaced in logarithmic scale and with approximately the same number of elements per bin: $M \ge$ 0.50\,M$_\odot$, 0.50\,M$_\odot > M \ge$ 0.25\,M$_\odot$, 0.25\,M$_\odot > M \ge 0.10$\,M$_\odot$, and $M < 0.1$\,M$_\odot$.
The uppermost and lowermost bins go up to 3.6\,M$_\odot$ and down to about the deuterium burning mass limit at 0.013\,M$_\odot$, respectively.
Finally, we counted the number of single, double, triple, quadruple, and quintuple systems in each mass bin, and determined the MF and CSF.
Table~\ref{tab:MF_CSF} summarises our results.
As expected, both MF and CSF decrease as stellar mass decreases \citep[][and references therein]{duchene13}.
Because of the completeness and comprehensiveness of our analysis, these MF and CSF values are among the most accurate ones determined to date for stars less massive than the Sun.
However, a larger, deeper survey is needed to increase the precision, and hence, completeness at the faintest end, as well as to quantify the rate of high-order multiples at the bottom of the main sequence, and link it to very low-mass star formation scenarios.

\subsection{Magnitudes, separations, masses, binding energies, and orbital periods}
\label{sec:parameters}

\begin{figure}
    \centering
    \includegraphics[width=1\linewidth]{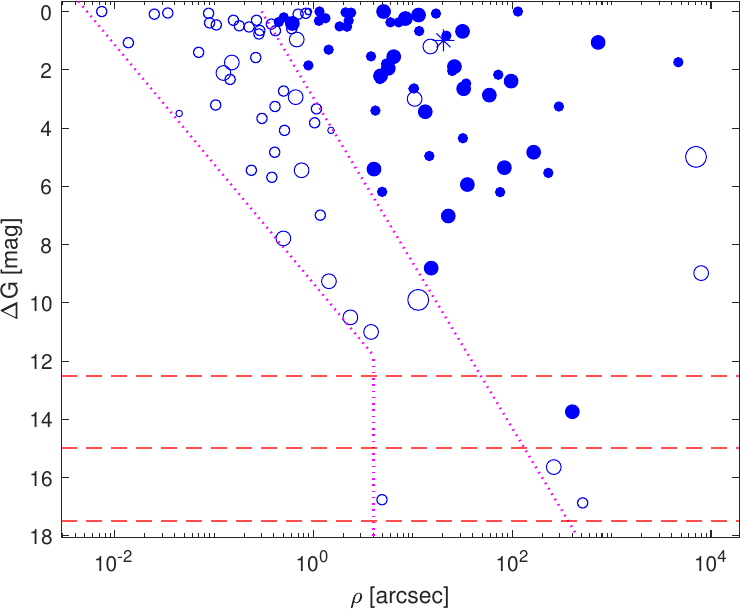}
    \caption{Absolute magnitude difference in $G$ band as a function of angular separation for each resolved pair. 
    Filled circles represent pairs with measured magnitudes of both components, while empty circles represent pairs with estimated magnitudes for one or both components. 
    Symbol size is approximately proportional to the mass of the primary (the only double white-dwarf binary is marked with an asterisk).
    Dashed horizontal lines indicate primary-to-companion contrast ratios of $10^5$, $10^6$, and $10^7$ from top to bottom.
    The diagonal and vertical lines indicate the empirical inaccessible regions with current facilities (e.g. \textit{Gaia}, adaptive optics at 2--8\,m-class telescopes, speckle, and lucky imaging) for photometry of both pair components (right) and only one component (left).
    The four companions with contrast greater than $10^5$ are T dwarfs close to the bright primaries $\epsilon$~Ind, $\xi$~UMa, HD~425818, and HD~131977 (GJ~570).} 
    \label{fig:10pc_rhoG}
\end{figure}

We reviewed the relation between magnitudes and angular separations of the investigated pairs.
Fig.~\ref{fig:10pc_rhoG} shows the magnitude difference in the \textit{Gaia} $G$ band between components in resolved pairs as a function of their angular separation, $\rho$ (or $\alpha$ when known).
As expected, there is a lack of very close pairs with high contrast ratios between components.
This contrast is independent of the primary spectral type.
At angular separations greater than 1\,arcsec, \textit{Gaia} is mainly constrained by its magnitude limit at $G \approx 20.4$\,mag, except for naked-eye primaries.
At shorter angular separations, \textit{Gaia} is able to measure magnitudes of two equal-brightness components separated by $\sim$0.4\,arcsec \citep[][their fig.~12 -- although \citealt{ziegler18} and \citealt{brandeker19} provided slightly larger values]{cifuentes25}, while some ground facilities, especially interferometers, can reach down to $\sim$0.01\,arcsec.

Taking advantage of our sample, we also addressed a problem of very close pairs with blended \textit{Gaia} EDR3 photometry that was proposed by \citet{golovin23} when studying their Fifth Catalogue of Nearby Stars (CNS5).
Because of contamination by close secondaries, primary stars suffering from blending have redder $G-RP$ colours than single stars of identical spectral type and metallicity.
This effect does not arise from simple aperture differences in $BP$ (and $RP$) in $G$, but from the nature of \textit{Gaia} photometry. 
While the $G$ magnitude is derived from one-dimensional profile-fitting photometry (line spread function fitting along the along-scan direction on CCD windows; \citealt{gaiacollaboration21a}), the $BP$ and $RP$ bands are more affected by blending and extended flux. 
As a result, unresolved doubles show artificially reddened colours.
In our sample, this blending (or artificial reddening) indeed occurs with \textit{Gaia} DR3 photometry for pairs separated by less than 2.2\,arcsec (and with secondaries brighter than the \textit{Gaia} magnitude limit). 
This fact highlights the need to empirically derive corrected $G - RP$ colours to settle photometric inconsistencies.

\begin{figure}
    \centering
    \includegraphics[width=1\linewidth]{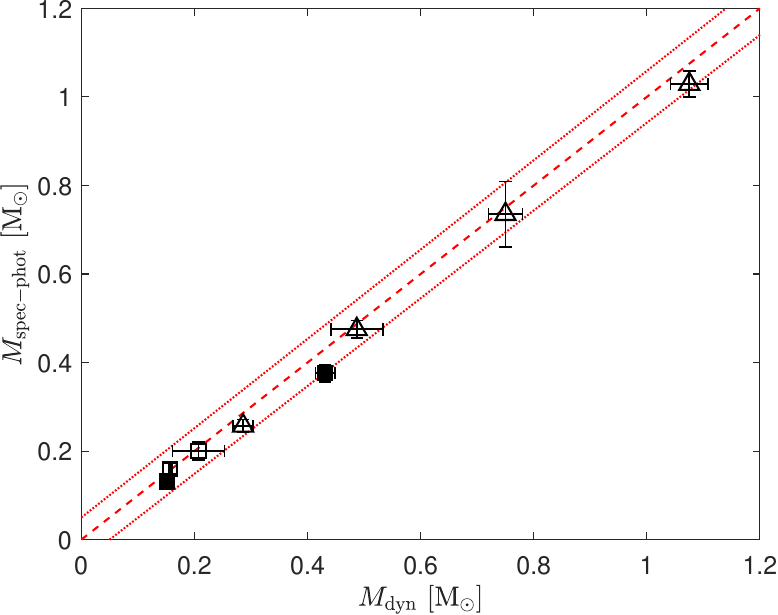}
    \caption{Comparison between dynamical masses $M_{\rm dyn}$ computed by us and spectro-photometric masses $M_{\rm spec-phot}$ determined by \citet{kirkpatrick24}. 
    Triangles and squares denote `primary' and `secondary stars, respectively, of the systems with {\tt orvara} orbital solutions in Table~\ref{tab:new_orbital_solutions}, while filled symbols indicate close pairs with combined masses (BD--17~588[BC] and BD+16~2708B[a,b]). 
    The dashed line marks the 1:1 ratio, with $\pm$5\% offsets marked with dotted lines.}
    \label{fig:masses}
\end{figure}

The earliest and, therefore, most massive stars in multiple systems within 10\,pc are Sirius~[A] (A0\,V), Fomalhaut (A4\,V), Procyon~A (F5\,IV--V), and $\gamma$~Lep (F6\,V), all with well-determined masses greater than 1.2\,M$_\odot$.
Conversely, there are 13 T dwarfs and one early Y dwarf in binary (WISE J045853.89+643452.5 [AB], Scholz’s Star [B], Luhman~16\,B, WISE J121756.90+162640.8 [AB], ULAS J141623.94+134836.3, SCR J1845--6357\,B), triple (GJ~229\,B[ab], $\epsilon$~Ind B[ab]), quadruple (GJ~570\,D), and quintuple systems ($\xi$~UMa~C).
In all cases except for GJ~229\,B[ab], their masses between the hydrogen and deuterium mass limits are poorly constrained \citep{caballero18}.
The L-dwarf population is a mixture of very low-mass stars and brown dwarfs with masses below the substellar boundary, and their masses are subject of uncertainty due to their unknown age. 
However, the masses of GKM stars are relatively accurate and some of them have been measured dynamically.
In particular, as illustrated by Fig.\,\ref{fig:masses}, the dynamical masses for the components of four wide systems that determined (Table\,\ref{tab:new_orbital_solutions}) fit very well previous spectro-photometric determinations.

With the masses and corresponding (projected) physical separations, we computed reduced binding energies, $|U_g^*|$, for each system.
The values of $|U_g^*|$ of all systems are greater than the minimum value required to avoid disruption by the galactic potential \citep{bahcall81,weinberg87,jiang10}, as well as the lowest reduced binding energy observed in moderately separated, very low-mass binary systems \citep{chauvin04,artigau07,caballero07a,radigan09}, which is about 10$^{33}$\,J \citep{caballero10}. 

Angular separations, $\rho$, of resolved systems range from less than 10\,mas (G~184--19~AC, GJ~229\,B[ab], EZ~Aqr~BC), to more than 1000\,arcsec (AU~Mic, Fomalhaut, and $\alpha$~Cen systems), and corresponding projected physical separations from less than 0.1\,au to over 100\,000\,au.
Likewise, the compiled and estimated orbital periods range more than ten orders of magnitude from $P = 1.795 \pm 0.017$\,d for the single-lined spectroscopic binary FL~Aqr~[AB] \citep{davison14} to tens of million years for AU~Mic and AT~Mic and Fomalhaut [A] and C, which reach the blurred boundary between the widest multiple systems and chance alignments in young stellar kinematic groups \citep{caballero10,shaya11,mamajek13}. 

\subsection{The closest neighbours}
\label{sec:neighbours}

Finally, we compared the separations of the widest multiple systems with the distances between the closest, second-closest, and third-closest neighbours (either single or primary stars in multiple systems).
On one hand, there are only three systems with projected physical separations greater than 10$^4$\,au ($s \sim$ 0.0485\,pc), namely $\alpha$~Cen~AB and Proxima ($s \sim$ 0.053\,pc), AU~Mic and AT~Mic~AB ($s \sim$ 0.22\,pc), and 
Fomalhaut A, B, and C ($s \sim$ 0.26--0.76\,pc)\footnote{Besides, there are eight companions at between $10^3$\,au and $10^4$\,au to their primaries, namely Wolf~47, BX~Cet, $\xi$~UMa~C, GJ~570\,D, vB~8, V2215~Oph, Ross~730, and $\epsilon$~Ind~Bab.}.
On the other hand, the mean distances between the closest, second-closest, and third-closest neighbours are 1.2\,pc, 1.7\,pc, and 2.1\,pc, respectively.
The inner tail of the distribution of distances between the closest neighbours extends well below 0.5\,pc (there are nine such pairs).
In particular, there are three neighbours with separations of 0.18--0.29\,pc (namely L~32--9 and SCR~J0630--7643AB, L~737--9 and LP~776--46, and 20~Crt~AB and WISE J114156.71--332635.8).
However, none of them satisfy our common-proper-motion criterion.
As a result, the tails of the distributions of the separations of the widest multiple systems and of the closest (in space) non-related stars slightly overlap between 0.18\,pc and 0.76\,pc, but a simple proper-motion filter is enough for disentangling between the two populations.
This overlapping region corresponds to orbital periods of tens of millions years, where our log-normal fit flattens.
AU~Mic and AT~Mic~AB and Fomalhaut A, B, and C, which are members of very young stellar kinematic groups \citep{luyten1938,mamajek13,alonsofloriano15b,riedel17,gagne18,gonzalezpayo23,cortescontreras24}, may not be a challenge to the log-normal distribution of binaries in the solar neighbourhood, but actual physically-bound systems caught in the process of disintegration by the galactic gravitational potential \citep{caballero09,shannon14}.

\section{Summary and future work}
\label{sec:summary}

In this study, we analysed the multiplicity of all stars and brown dwarfs at a maximum heliocentric distance of 10\,pc, originally compiled by \citet{reyle22}. 
Our analysis focused on identifying, confirming, and characterising the multiple systems in the sample, compiling and determining their astrometric properties and masses, and deriving their orbital periods and reduced binding energies. 
For this purpose, we used data from \textit{Gaia} DR3 for a common parallax and proper motion search and the WDS and MCS catalogues, together with available literature.
For seven resolved systems, we determined new orbital solutions from public astrometric and radial-velocity data and computed periods ranging from 14.1 to 591\,a.
As a result, we identified 215 stars and brown dwarfs in 92 systems at all physical separations from less than 0.1\,au to near half a parsec. 
Of these, 68 systems are double, 19 are triple, 3 are quadruple, and 2 are quintuple.
Only eight systems contain at least one unresolved pair.
In terms of the mass distribution of stars in multiple systems, there are four stars with masses over 1.2\,M$_\odot$, and 13 T dwarfs and one early Y binary at the opposite mass end, well below the hydrogen burning mass limit.
We measured the multiple and companion star fractions of the sample, MF$_{\rm all} = 26.2^{+4.8}_{-4.3}$\,\% and CSF$_{\rm all} = 0.350^{+0.051}_{-0.048}$, and also of M dwarfs only and of four different mass ranges, which confirm the observed trend of lower MF and CSF for lower masses of primary stars.
The computed (reduced) binding energies indicate that all systems are stable enough to survive over time against disruptive galactic forces.
Finally, we compiled, estimated, or determined orbital periods for all the 123 identified pairs, ranging from 1.795\,d to tens of million years. 
Lastly, the shortest distances among stars in our neighbourhood slightly overlap with the largest separated, weakly bound systems that are being  
disrupted by galactic tidal forces.

After this comprehensive data compilation and analysis, still dozens of systems need further investigation. 
On one hand, poorly investigated systems require accurate relative astrometry and precise radial velocities for determining orbital solutions and, therefore, dynamical masses, periods, and semimajor axes.
On the other hand, better studied systems with white-dwarf, ultracool-dwarf, and exoplanet companions need further investigation, such as precise magnitude differences and spectral types of close resolved companions, stellar and planet mass improvement, ratio of orbital separations in multiple systems, or orbits in 3D.
Furthermore, there can be missing multiple stellar systems within 10\,pc, yet to be discovered -- albeit very few. 
However, despite the extraordinary amount of data currently available, the parameter space of very close ($\rho<$~1\,au), low-mass, and high-contrast companions remains only partially explored, leaving room for additional detections, especially at very low masses in the brown dwarf and planetary regimes.
More future work is expected, though, to extend this sample to 20\,pc and beyond, and quantify its completeness.

Data from \textit{Gaia} DR4, to be released on 2 December 2026, will certainly help in better understanding our closest multiple systems.
We underline the significant impact that the next \textit{Gaia} release will have for samples in the solar neighbourhood: Apart from cataloguing millions of new non-single star solutions, it will publish astrometric and radial-velocity epoch observations for the 66 months between July 2014 and January 2020, which when combined with ground-based observations spread over decades, will reveal new unresolved systems.
However, there must be a concerted effort worldwide to characterise them in detail beyond \textit{Gaia}'s capabilities.
Actually, some of the systems reported here are in the lists of potential targets of HWO and LIFE, so this work also does its part in the search for habitable exoplanets.

To sum up, this paper provides an updated census of single stars, binary, and multiple systems in the close solar
vicinity ($d \lesssim$ 10\,pc). 
Although the volume is limited and the resulting sample is small, we strove for a complete and unbiased probe of the stellar multiplicity phenomenon and capitalised on the amount of collected information about these historically best studied objects.

\section*{Acknowledgements}
We thank the anonymous reviewer, S.~Deveny, E.~Gonz\'alez-\'Alvarez, Z.~Hartman, S.\,B.~Howell, S.~Ilovaiski, C.~Littlefield, E.\,E.~Mamajek, R.~Matson, J.~Sahlmann, R.\,L.~Smart, C.~Reyl\'e, A.~Tokovinin, and M.\,R.~Zapatero Osorio for providing numerous helpful comments and data.
We acknowledge financial support from the Agencia Estatal de Investigaci\'on (AEI/10.13039/501100011033) of the Ministerio de Ciencia e Innovaci\'on and the European Regional Development Fund ``A way of making Europe'' through projects
PID2022-137241NB-C4[2,4], 
PID2023-150468NB-I00,   
and PID2021-125627OB-C31. 
We also acknowledge financial support by the Consejo Superior de Investigaciones Cient\'ificas through internal project 2023AT003 associated to RYC2021-031640-I.
This research made use of the NASA's Astrophysics Data System Bibliographic Services,
the Exoplanet Archive, 
the Extrasolar Planets Encyclopaedia,
the Washington Double Star catalogue maintained at the U.S. Naval Observatory, 
and the Simbad database \citep{wenger00}, 
VizieR catalogue access tool \citep{ochsenbein00}, 
and Aladin sky atlas \citep{bonnarel00} of the Centre de Donn\'ees astronomiques de Strasbourg (France).

\section*{Data availability}
\label{sec:data_availability_10pc}

Tables~\ref{tab:sample_10pc}, \ref{tab:systems_10pc}, and \ref{tab:pairs_10pc} are available at the CDS via anonymous ftp to \texttt{cdsarc.cds.unistra.fr} (130.79.128.5) or via \texttt{https://cdsarc.cds.unistra.fr/viz-bin/cat/J/MNRAS/}.


\bibliographystyle{mnras}
\bibliography{biblio}


\appendix

\section{Remarkable systems}
\label{sec:remarkable_systems_10pc}

\textbf{WDS~J00363+1821.}
LSPM~J0036+1821 (2MASS J00361617+1821104) was first discovered by \citet{reid00} and later classified by \cite{reid06} as an L3.5+L6: unresolved ultra-cool dwarf pair. 
The system was subsequently resolved by \citet{bernat10} with aperture masking interferometry and Palomar laser-guide-star adaptive optics in the $K_s$ band.
The reported angular separation was only $\rho=$ 89.5\,mas.
This discovery was later confirmed by \cite{pope13} in the $J$ and $H$ bands with a kernel phase interferometry reanalysis of archival NICMOS/{\it Hubble}.
There are new unpublished data that support its binarity (P.~Miles-P\'aez, priv.~comm.).

\smallskip
\noindent\textbf{WDS~J06106--2152.}
HD~42581 (GJ\,229) is a triple system consisting of an M dwarf and two brown dwarfs.
Considered single for several decades, GJ\,229\,B was the first T dwarf to be detected \citep{nakajima95} with confirmed methane absorption features \citep{oppenheimer95, oppenheimer98}.
It orbits the primary at $a=$ 33.3\,au with a period of 237.9\,a \citep{brandt21b}.
The T dwarf was further resolved into a 12.1-day binary with $a=$ 0.042\,au \citep{xuan24}. 
Additionally, two controversial planetary companions to GJ\,229\,A have been proposed (\citealt{tuomi14,feng20,feng22}).

\smallskip
\noindent
\textbf{WDS~J06523--0510.}
HD~50281~AB consists of an early-K primary and a early-M secondary.
The pair has been known since the middle of the 20th century \citep[e.g.][]{eggen56}.
WDS tabulates two more pairs in the system: 
A 9.6\,arcsec companion candidate reported by \citet{tanner10} is a background star according to \textit{Gaia} DR3, while the hypothetical close binarity of the secondary (WSI~BaBb, $\rho = 0.04$\,arcsec on a single epoch in 2010, $\Delta V \sim$ 0.5\,mag) has been repeatedly questioned \citep{tokovinin15, tokovinin20, tokovinin18, mason18}. 

\smallskip
\noindent
\textbf{WDS~J08589+0829.}
G~41--14 \citep{giclas59} is a triple system made of roughly similar M-type dwarfs \citep{stephenson86,kirkpatrick12}.
The primary is a 7.6\,d-period double-lined spectroscopic binary \citep{reid97b}, while the third component was discovered by \citet{delfosse99} with adaptive optics at an angular separation of 0.620\,arcsec.
They estimated an orbital period of approximately 10 years. 
After two decades collecting more astrometric data \cite[e.g.][]{hartkopf12, janson14}, the actual orbital period is about half that value \citep{tokovinin23}.

\smallskip
\noindent
\textbf{WDS~J10509+0648.}  
\citet{lamman20} reported a close companion to Wolf~358 (EE~Leo) with a separation of 0.098$\pm$0.070\,arcsec, detected in a single Robo-AO observation from 2016.
No further confirmation of multiplicity has been reported. 
In fact, it appears to be single based on extremely precise astrometry from \textit{Gaia} \citep[e.g.,][]{kervella19, vrijmoet20} and, in particular, radial velocity measurements from CARMENES \citep{ribas23}.  

\smallskip
\noindent
\textbf{WDS~J11182+3132.}
$\xi$~UMa (Alula~Australis) is a quintuple system.
According to \citet{mason95}, it was one of the first binaries reported \citep{herschel1803}, with precise relative astrometry ($\Sigma$~1523: \citealt{struve1827}), calculated orbit \citep{savary1827}, and also a ``definitive'' orbit \citep{norlund1905,vandenbos1928,heintz67,worley83}.  
The two main components of $\xi$~UMa are main-sequence, solar-like stars with angular separation of about 1–3\,arcsec, with an orbital period of 59.89\,a. 
Each component of this double star is itself a close binary. 
The 1.835\,a eccentric orbit of the A pair has been determined from spectroscopy and speckle interferometry \citep{heintz67,mason95}, while $\xi$~UMa~B is a single-lined spectroscopic binary with a circular 3.98-day orbit \citep{berman31,griffin98}. 
Using Wide-field Infrared Survey Explorer (WISE) data, \citet{wright13} discovered the fifth component of the system, WISE~J111838.70+312537.9 ($\xi$~UMa~C), a T8.5 brown dwarf located 8.5\,arcmin away.

\smallskip
\noindent
\textbf{WDS~14545+1606.}
BD+16~2708 (CE~Boo, GJ~569) is a triple system consisting of a M3.0\,V primary and a close pair of ultracool dwarfs of spectral type M8.5--9.0\,V in a hierarchical arrangement ($\rho_{\rm A-Ba,Bb} \sim$ 5.0\,arcsec, $\rho_{\rm Ba-Bb} \sim$ 0.1\,arcsec). 
The system has been widely investigated because it may contain at least one brown dwarf \citep{forrest88,martin00,zapateroosorio04,simon06,konopacky10,femenia11,dupuy17}.
In spite of previous analyses, we determined for the first time the orbit parameters of the A--Ba,Bb pair, with a period of $450^{+52}_{-39}$\,a, and the dynamical masses of the wide pair components.
On one hand, the primary has a dynamical mass $M_{\rm A} = 0.493^{+0.046}_{-0.039}$\,M$_\odot$, slightly greater than, but identical within uncertainties to, masses estimated by \citet{kirkpatrick24} and \citet{cifuentes25} from spectro-photometry at $0.475\pm0.020$\,M$_\odot$ and $0.477\pm0.033$\,M$_\odot$, respectively.
On the other hand, from our simultaneous fit with {\tt orvara} of astrometric and radial-velocity data, we measured a combined mass of the close pair $M_{\rm Ba,Bb} = 0.1503\pm0.0035$\,M$_\odot$, slightly above the combined masses in previous work.
Using the conservative mass ratio of $q = 0.80\pm0.15$ of \citet{femenia11} for the B components, we estimated masses $M_{\rm Ba} = 0.084\pm0.016$\,M$_\odot$ and $M_{\rm Bb} = 0.067\pm0.013$\,M$_\odot$.
The secondary has therefore a most probable mass below the hydrogen burning limit.
With a moderately young age of 200--500\,Ma and a distance of about 10\,pc, this system deserves further investigation.

\smallskip
\noindent
\textbf{WDS~J16555--0820.}
This complex hierarchical system \citep[or mini-cluster, e.g.][]{eggen65} is quintuple and the closest such system to the Sun \citep{wolf1917,kuiper34}.
The primary component, HD~152751, has an M3.0\,V spectral type \citep{reid95} and is a triple-lined spectroscopic binary \citep{joy47,delfosse99}, with inner and outer orbital periods of 2.965 and 627 days, respectively, and a combined mass of roughly one solar mass \citep{mazeh01}. 
Later, \citet{mason23} was able to resolve the long-period pair.
Wolf~629, an M3.5\,V star \citep{reid95}, is the fourth component, located 72.3\,arcsec away from the primary.  
The fifth component, vB~8, is an M7.0\,V star located 231\,arcsec from the primary \citep{vanbiesbroeck61,alonsofloriano15a}. 
Early infrared speckle interferometry by \citet{mccarthy85} suggested the presence of a cool companion at 1\,arcsec to vB~8, consistent with a brown dwarf. 
However, this claim was later disproven and vB~8 was confirmed as a single star within the AstraLux sensitivity range \citep{janson14}.  

\smallskip
\noindent
\textbf{G~203--47.}
This is a single-lined spectroscopic binary discovered by \citet{reid97b}, composed by an M3.5\,V star and a white dwarf, with the first (and only) orbit determination by \citet{delfosse99}. 
A large radial velocity scatter has been observed in several surveys, including \textit{Gaia} DR3, but no re-determination of the orbit has been made since.

\smallskip
\noindent
\textbf{WDS~J22388--2037.}
FK~Aqr and FL~Aqr form a visual pair ($\rho \approx 25$\,arcsec) of magnetically-active cool dwarfs \citep[][and references therein]{vyssotsky52,herbig65,tsvetkova24}.
Both components have been further identified as close binaries, therefore constituting a quadruple star system.
FK~Aqr is a double-lined spectroscopic binary in a 4.083-day orbit with astrometric signatures consisting of two early-M dwarfs of nearly equal mass \citep{herbig65,delfosse99,kervella19}.
FL~Aqr is a single-lined spectroscopic binary with an ultra-cool dwarf companion in a 1.795-day orbit \citep[][who reported a minimum mass of the companion of 0.056\,M$_{\odot}$]{davison14}.
This is the only quadruple system within 10\,pc of the Sun comprising just M-type (or M- and L-type) components.

\clearpage
\onecolumn

\section{Plots}
\label{sec:plots_10pc}

\begin{figure}
 \begin{subfigure}{.48\textwidth}
  \centering
  \includegraphics[width=1\linewidth, angle=0]{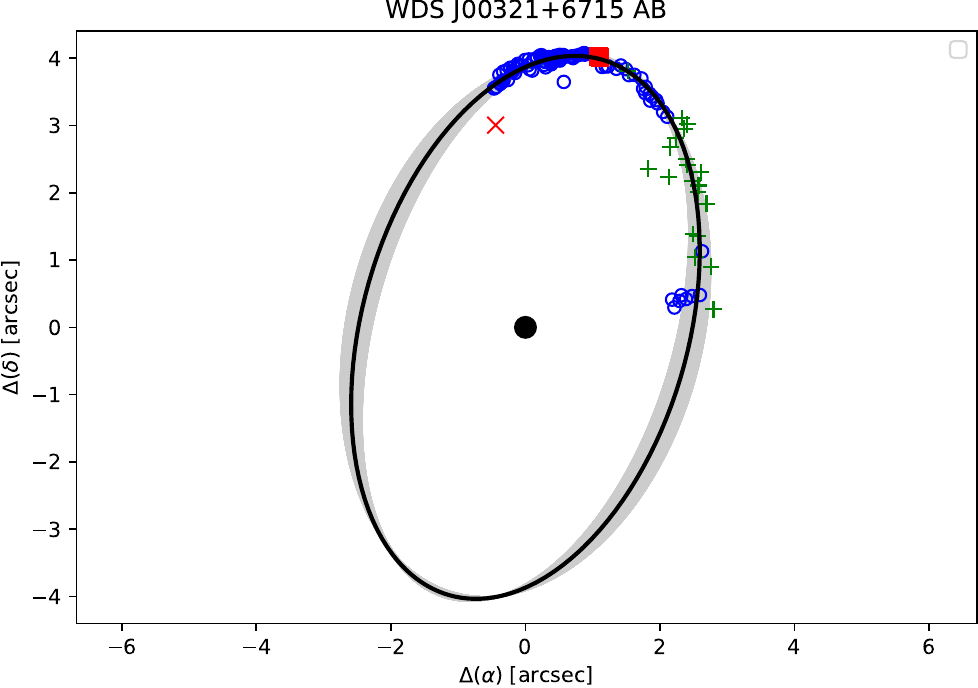}
 \end{subfigure}
  \begin{subfigure}{.48\textwidth}
  \centering
  \includegraphics[width=1\linewidth, angle=0]{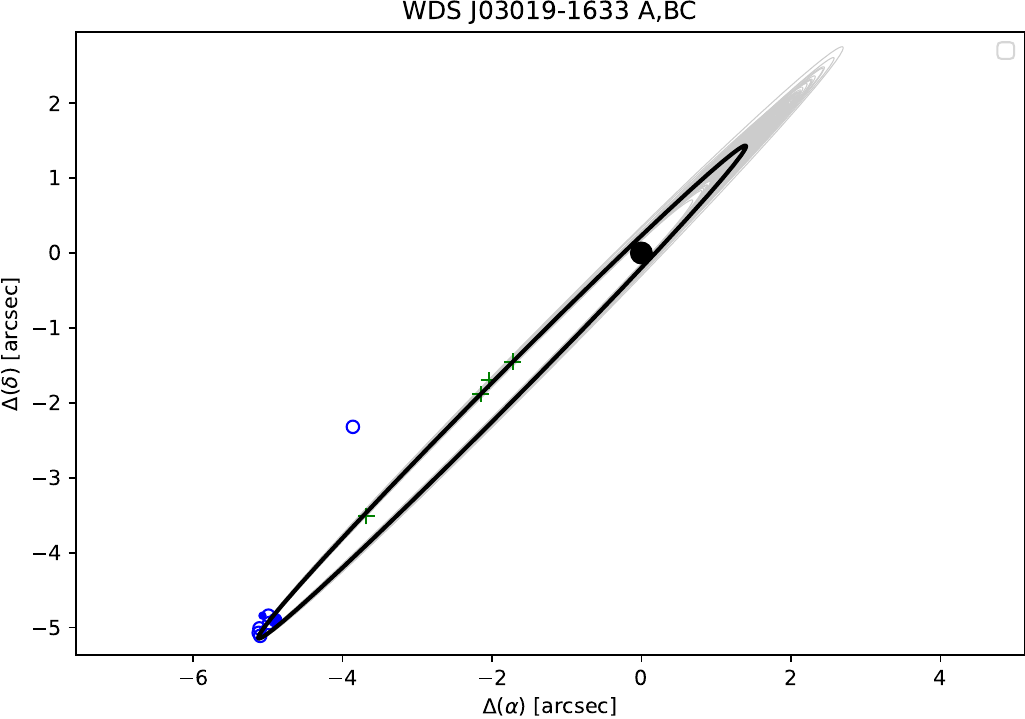}
 \end{subfigure} 

 \vspace{5mm}
 \begin{subfigure}{.48\textwidth}
  \centering
  \includegraphics[width=1\linewidth, angle=0]{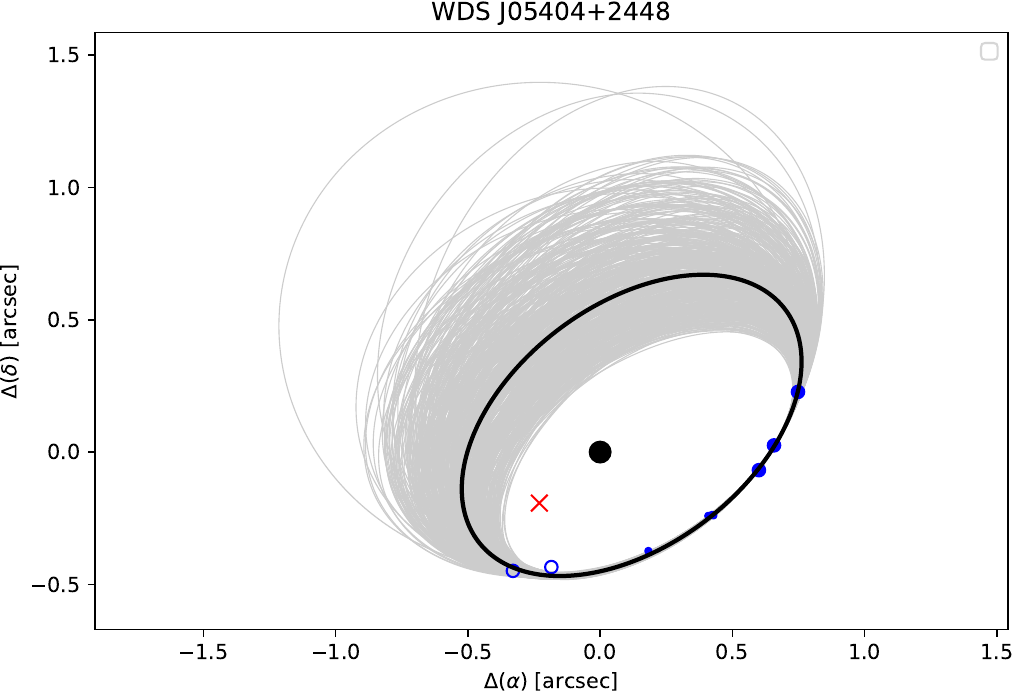}
 \end{subfigure}
  \begin{subfigure}{.48\textwidth}
  \centering
  \includegraphics[width=1\linewidth, angle=0]{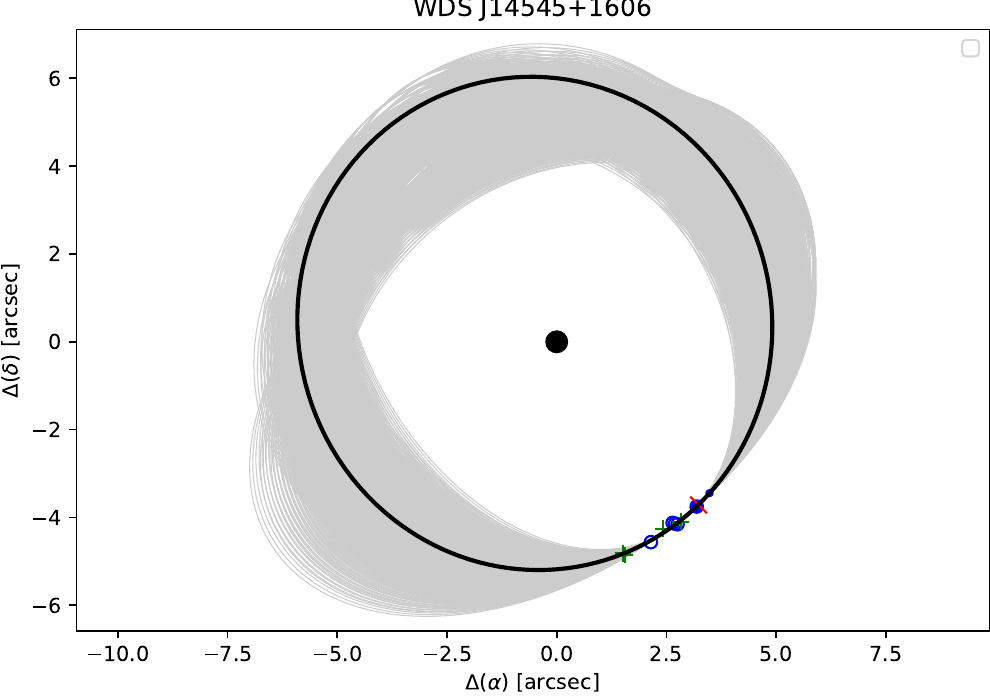}
 \end{subfigure} 

 \vspace{5mm}
 \begin{subfigure}{.48\textwidth}
  \centering
  \includegraphics[width=1\linewidth, angle=0]{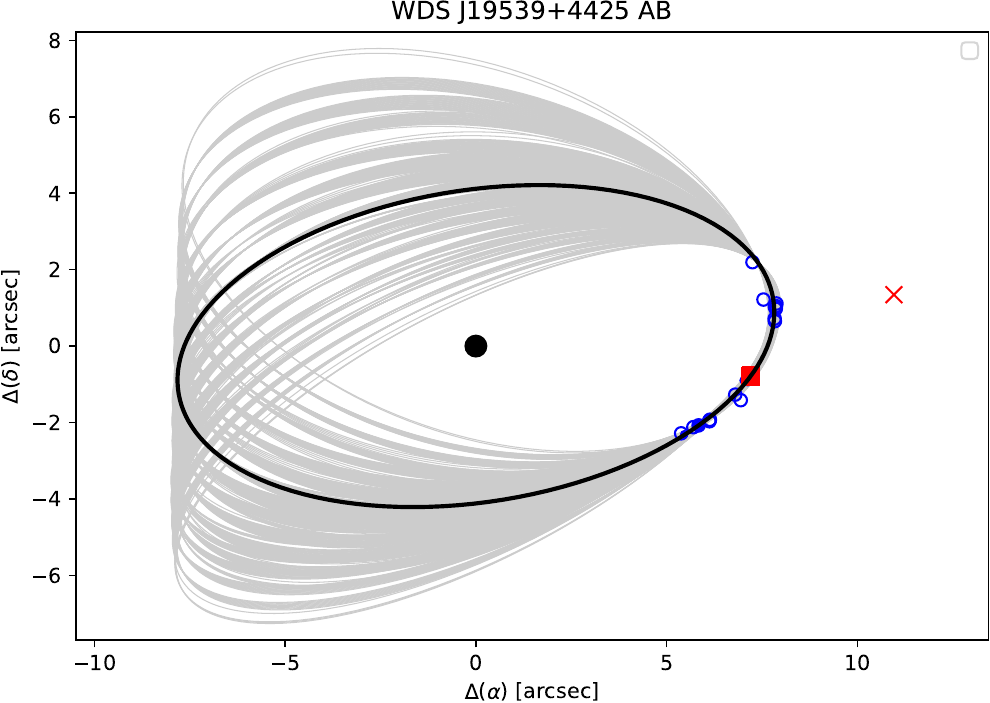}
 \end{subfigure}
  \begin{subfigure}{.48\textwidth}
  \centering
  \includegraphics[width=1\linewidth, angle=0]{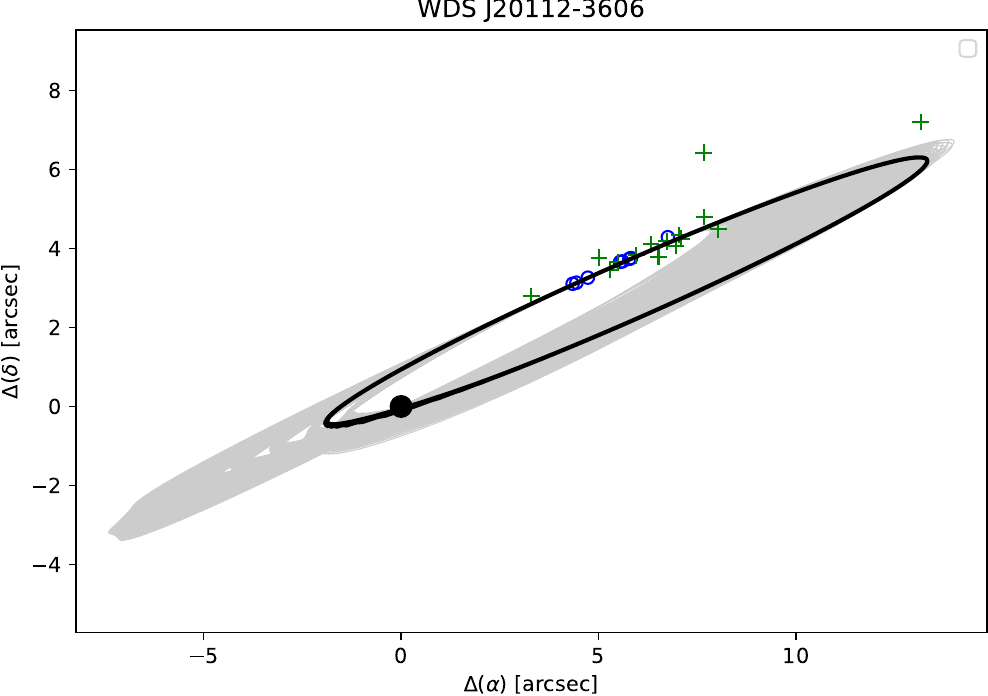}
 \end{subfigure} 

 \caption{New orbital solutions for six wide pairs in Table~\ref{tab:new_orbital_solutions} ($\chi^{01}$~Ori is in Fig.~\ref{fig:chi01_Ori}).
 The black filled circle is the primary, while the 
 solid blue circles (speckle interferometry),
 open blue circles (CCD and photographic measures),
 solid red squares (\textit{Hipparcos/Tycho-2} astrometry), green pluses (micrometric measures), and red crosses (rejected measures) indicate the relative astrometric measurements of the companion.
The light gray lines represent orbital realisations within the confidence region associated with the best-fitting solution (between 118 and 500 per plot).
 North is to the bottom, East is to the right.}
\label{fig:10pc_orbital_solutions}
\end{figure}

\newpage
\onecolumn

\section{Long tables}
\label{sec:sample_of_work}

\footnotesize

\begin{longtable}{llccccc}
\caption{Stars and brown dwarfs at $d <$ 10\,pc.}\\
\\
\noalign{\smallskip}
\hline 
\noalign{\smallskip}
Star or brown dwarf & GJ$^{(a)}$ & $\alpha$ (J2000) & $\delta$ (J2000) & $\mu_{\alpha}\cos\delta$ & $\mu_\delta$ & $d$ \\
 &  & (hh:mm:ss.ss) & (dd:mm:ss.s) & (mas\,a$^{-1}$) & (mas\,a$^{-1}$) & (pc)  \\
\noalign{\smallskip}
\hline
\noalign{\smallskip}
\endfirsthead
\caption{Stars and brown dwarfs at $d <$ 10\,pc \textit{(continued).}}\\
\\
\hline 
\noalign{\smallskip} 
Star or brown dwarf & GJ$^{\text{(a)}}$ & $\alpha$ (J2000) & $\delta$ (J2000) & $\mu_{\alpha}\cos\delta$ & $\mu_\delta$ & $d$ \\
 &  & (hh:mm:ss.ss) & (dd:mm:ss.s) & (mas\,a$^{-1}$) & (mas\,a$^{-1}$) & (pc)  \\
\noalign{\smallskip}
\hline
\noalign{\smallskip}
\endhead
\hline
\noalign{\smallskip}
\endfoot 

LAWD 96 & 915 & 00:02:10.72 & --43:09:55.4 & 613.79 & --686.99 & 8.33 \\
WISE J000517.48+373720.5 & -- & 00:05:17.48 & +37:37:20.5 & 997.30 & --271.60 & 7.88 \\
HD 225213 & 1 & 00:05:24.43 & --37:21:26.5 & 5633.44 & --2334.72 & 4.35 \\
G 158--27 & 1002 & 00:06:43.20 & --07:32:17.0 & --811.57 & --1893.25 & 4.85 \\
GAT 1367 & -- & 00:11:31.82 & +59:08:39.9 & --905.72 & --1166.83 & 9.31 \\
G 158--50 [AB] & 1005 AB & 00:15:28.77 & --16:08:11.7 & 731.83 & --607.73 & 6.00 \\
HD 1326 [A] & 15 A & 00:18:22.88 & +44:01:22.6 & 2891.52 & 411.83 & 3.56 \\
HD 1326B & 15 B & 00:18:25.83 & +44:01:38.1 & 2862.80 & 336.43 & 3.56 \\
\end{longtable}
\begin{justify}
   \footnotesize{\textbf{\textit{Notes. }}}
   \footnotesize{$^{\text{(a)}}$ For homogeneity and readability, we write `GJ' \citep{gliese79,gliese91} even for stars in the catalogue of nearby stars of \citet{gliese69}, which originally had a `Gl' identifier. We refrain from tabulating stars with `GJ' identifier $\ge$ 9000, as they were not part of any Gliese or Gliese-Jahreiss catalogues.}
\end{justify}   

\newpage
\pagebreak[4]
  \global\pdfpageattr\expandafter{\the\pdfpageattr/Rotate 90}

\begin{landscape}
\scriptsize
\begin{longtable}{lllllccccccllcc}
\caption{Individual components in all multiple systems at $d <$ 10\,pc.}\\
\label{tab:systems_10pc}\\
\noalign{\smallskip}
\hline
\noalign{\smallskip}
WDS & Star or brown dwarf & Alternative & GJ & Comp.$^{(\text{a})}$ & $\alpha$ & $\delta$ & $d$ & Spectral & $G^{(\text{b})}$ & $M$ & Ref. & Planets$^{(\text{d})}$ & $|U^*_g|$ & System \\
 & name & name & & & [J2000] & [J2000] & (pc) & type$^{(\text{b})}$ & (mag) & (M$_\odot$) & $M^{(\text{c})}$ &  & (10$^{33}$\,J) & diagram$^{(\text{e})}$ \\
\noalign{\smallskip}
\hline
\noalign{\smallskip}
\endfirsthead
\caption{Individual components in all multiple systems at $d <$ 10\,pc \textit{(continued).}}\\
\hline
\noalign{\smallskip}
WDS & Star or brown dwarf & Alternative & GJ & Comp.$^{(\text{a})}$ & $\alpha$ & $\delta$ & $d$ & Spectral & $G^{(\text{b})}$ & $M$ & Ref. & Planets$^{(\text{d})}$ & $|U^*_g|$ & System \\
 & name & name & & & [J2000] & [J2000] & (pc) & type$^{(\text{b})}$ & (mag) & (M$_\odot$) & $M^{(\text{c})}$ &  & (10$^{33}$\,J) & diagram$^{(\text{e})}$ \\
\noalign{\smallskip}
\endhead
\noalign{\smallskip}
\endfoot
  00155--1608 & G 158--50 [A] & -- & 1005 & [A] & 00 15 28.11 & --16 08 01.6 & 5.90 & M4.0\,V & 10.27 & 0.179$\pm$0.002 & Kir24 & -- &  & \multirow{2}{*}{\includegraphics[width=8.5mm]{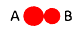}}\\
   & G 158--50 [B] & -- & -- & [B] & -- & -- & -- & M4.5:\,V & 10.7: & 0.112$\pm$0.001 & Kir24 & -- & 19\,000 & \\
\noalign{\smallskip}
\hline
\noalign{\smallskip}
  00184+4401 & HD 1326 [A] & Gmb 34, GX And & 15A & A & 00 18 22.88 & +44 01 22.6 & 3.56 & M1.0\,V & 7.22 & 0.395$\pm$0.012 & Cif25 & b c &  & \multirow{2}{*}{\includegraphics[width=5mm]{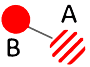}}\\
   & HD 1326B & GQ And & 15B & B & 00 18 25.83 & +44 01 38.1 & 3.56 & M3.5\,V & 9.69 & 0.164$\pm$0.010 & Cif25 & -- & 1200 & \\
\noalign{\smallskip}
\hline
\noalign{\smallskip}
  00247--2653 & LP 881--64 [A] & -- & 2005 & A & 00 24 44.18 & --27 08 25.3 & 7.73 & M5.5\,V & 13.16 & 0.112$\pm$0.001 & Kir24 & -- &  & \multirow{3}{*}{\includegraphics[width=6mm]{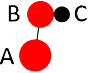}}\\
   & LP 881--64 [B] & -- & -- & B & -- & -- & -- & M9.5\,V & 16.5: & 0.084$\pm$0.001 & Kir24 & -- &  & \\
   & LP 881--64 [C] & -- & -- & C & -- & -- & -- & L0 & 16.8: & 0.080$\pm$0.001 & Kir24 & -- & 11\,000 & \\
\noalign{\smallskip}
\hline
\noalign{\smallskip}
  00321+6715 & BD+66 34A[a] & V547 Cas & 22A & Aa & 00 32 29.51 & +67 14 09.1 & 9.89 & M2.0 V & 9.57 & 0.405$\pm$0.008 & Kir24 & -- &  & \multirow{3}{*}{\includegraphics[width=8mm]{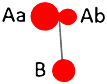}}\\
\end{longtable}
\begin{justify}
  \footnotesize{\textbf{\textit{Notes. }}}
   \footnotesize{$^{\text{(a)}}$ The letters to define components are shown in square brackets when they are not designated this way in the WDS catalogue.}
   \footnotesize{$^{(\text{b})}$ Colons indicate spectral type / $G$ magnitude estimations done by us.}
   \footnotesize{$^{(\text{c})}$ References:
      Bed24: \citet{bedin24};
      Bon07: \citet{bonavita07};
      Bur00: \citet{burgasser00};
      Cab09: \citet{caballero09};
      Cif25: \citet{cifuentes25};
      Dav14: \citet{davison14};      
      Del99: \citet{delfosse99};
      Fit24: \citet{fitzmaurice24};
      Fuh08: \citet{fuhrmann08};
      Gel11: \citet{gelino11};
      Ghe10: \citet{ghezzi10};
      Gon20: \citet{gonzalezalvarez20};     
      Kir24: \citet{kirkpatrick24};
      Liu12: \citet{liu12};
      Mal24: \citet{mallorquin24};
      Mar98: \citet{martin98};
      Maz01: \citet{mazeh01};      
      Mon06: \citet{montagnier06};
      Obr24: \citet{obrien24};
      Pop13: \citet{pope13};
      Tok25: \citet{tokovinin25};
      Vig12: \citet{vigan12}; 
      Win19: \citet{winters19};
      Wri13: \citet{wright13};
      Xua24: \citet{xuan24};
      Zha21: \citet{zhang21}.
     } 
     \footnotesize{$^{\text{(d)}}$ The letters that designate the planets are placed in brackets when their existence has not yet been confirmed.} 
     \footnotesize{$^{\text{(e)}}$ Colours in system diagrams indicate spectral types (dark blue: A, light blue: B, yellow: G, orange: K, red: black: M, L, T, and Y, white: WD). Concentric circles indicate unresolved binaries. 
     Stripped circles indicate exoplanet host stars. 
     Symbol sizes are approximate proportional to stellar masses, but are not to scale to the physical separations. 
     When separation is about 1\,arcsec or less, the black connection line is not drawn.}

 \end{justify}     
\end{landscape}

\begin{landscape}
\scriptsize
\begin{longtable}{lllllcccccccccl}
\caption{Pairs in all multiple systems at $d <$ 10\,pc.}\\
\label{tab:pairs_10pc}\\
\noalign{\smallskip}
\hline
\noalign{\smallskip}
WDS$^{\text{(a)}}$ & Disc. & Prim. & Sec. & Detection & $\theta$$^{\text{(d)}}$ & $\rho$ & $s$ & Ref. & $\alpha$$^{\text{(f)}}$ & $a$$^{\text{(g)}}$ & $P$ & $P$ & Orbit & Ref. \\
 & code & comp.$^{\text{(b)}}$ & comp.$^{\text{(b)}}$ & method$^{\text{(c)}}$ & (deg) & (arcsec) & (au) & $\theta,~\rho$$^{\text{(e)}}$ & (arcsec) & (au) &  & unit  & grade & $\alpha,~P$$^{\text{(e)}}$ \\
\noalign{\smallskip}
\hline
\noalign{\smallskip}
\endfirsthead
 \caption{Pairs in all multiple systems at $d <$ 10\,pc \textit{(continued).}}\\
\hline
\noalign{\smallskip}
WDS$^{\text{(a)}}$ & Disc. & Prim. & Sec. & Detection & $\theta$$^{\text{(d)}}$ & $\rho$ & $s$ & Ref. & $\alpha$$^{\text{(f)}}$ & $a$$^{\text{(g)}}$ & $P$ & $P$ & Orbit & Ref. \\
 & code & comp.$^{\text{(b)}}$ & comp.$^{\text{(b)}}$ & method$^{\text{(c)}}$ & (deg) & (arcsec) & (au) & $\theta,~\rho$$^{\text{(e)}}$ & (arcsec) & (au) &  & unit  & grade & $\alpha,~P$$^{\text{(e)}}$ \\
\noalign{\smallskip}
\endhead
\noalign{\smallskip}
\endfoot

00155--1608 & HEI 299 & [A] & [B] & HRI & -- & -- & -- & -- & 0.31285$\pm$0.00050 & \textit{1.845$\pm$0.011} & 4.55726$^{+0.00075}_{-0.00074}$ & a & 1.3 & Man19 \\
   \noalign{\smallskip}
\hline
\noalign{\smallskip}
00184+4401 & GRB 34 & A & B & \textit{Gaia} & -- & -- & -- & -- & \textit{26}$^{+\textit{12}}_{-\textit{1.7}}$ & 93$^{+42}_{-6}$ & 1.23$^{+0.93}_{-0.11}$ & ka & $<$5.4 & Pin18 \\
   \noalign{\smallskip}
\hline
\noalign{\smallskip}
00247--2653   & LEI 1 & A & BC & HRI & -- & -- & -- & -- & 1.593 & \textit{12.32} & 83.00 & a & $<$5.1 & MSC\\
   & LEI 1 & B & C & HRI & -- & -- & -- & -- & 0.460 & \textit{3.56} & 17.25 & a & $<$2.0 & MSC\\
   \noalign{\smallskip}
\hline
\noalign{\smallskip}
00321+6715 & MCY 1 & Aa & Ab & HRI & -- & -- & -- & -- & 0.51026$\pm$0.00074 & \textit{5.048$\pm$0.024} & 15.4275$\pm$0.0054 & a & 2.0 & Man19\\
   & VYS 2 & A & B & \textit{Gaia} & -- & -- & -- & -- & 4.138$^{+0.0056}_{-0.0058}$ & \textit{40.935$\pm$0.195} & 311$^{+26}_{-21}$ & a & $<$4.9 & This work\\
\noalign{\smallskip}
\hline
\noalign{\smallskip}
00363+1821   & BNA 1 & [A] & [B] & HRI & -- & 0.0445$\pm$0.0001 & 0.389 & Pop13 & -- & -- & 363.1$\pm$27.4 & d & -- & This work\\
   \noalign{\smallskip}
\hline
\noalign{\smallskip}
00491+5749   & STF  60 & A & B & \textit{Gaia} & -- & -- & -- & -- & 11.61$\pm$0.27 & \textit{68.8$\pm$1.6} & 453$\pm$15 & a & 2.9 & Izm19\\
\end{longtable}
\begin{justify}
   \footnotesize{
   \textbf{\textit{Notes. }}}
   \footnotesize{$^{\text{(a)}}$ When the WDS identifier is missing, the common system name is shown in parentheses.}   
   \footnotesize{$^{\text{(b)}}$ The `primary' and `secondary' components are shown in parentheses when they are not designated this way in the WDS catalogue.}
   \footnotesize{$^{\text{(c)}}$ DI: Direct imaging; \textit{Gaia}: \textit{Gaia} absolute astrometry; HRI: High resolution imaging (AO, speckle, \textit{Hubble}, interferometry, etc.); SB1/SB2: Single-/double-lined spectroscopic binarity.}
   \footnotesize{$^{\text{(d)}}$ If $\rho<$1\,arcsec, $\theta$ is not shown.}
   \footnotesize{$^{\text{(e)}}$ References:
      Ake21: \citet{akeson21}; 
      Bar12: \citet{barry12};
      Bed24: \citet{bedin24};
      Beu04: \citet{beuzit04};
      Bil06: \citet{biller06};
      Bon17: \citet{bond17};
      Bon20: \citet{bond20};      
      Bra21: \citet{brandt21b};
      Cal24: \citet{calamari24};
      Car09: \citet{cardoso09};
      Cur22: \citet{curiel22};
      Dav14: \citet{davison14};      
      Del99: \citet{delfosse99};
      Dup17: \citet{dupuy17};      
      Dup19: \citet{dupuy19};
      Far10: \citet{farrington10};
      Fen21: \citet{feng21};
      Fen22: \citet{feng22};
      Fit24: \citet{fitzmaurice24};
      Gon20: \citet{gonzalezalvarez20};
      Gra24: \citet{gravitycollaboration24};
      Gri98: \citet{griffin98};
      Hei87: \citet{heintz87};
      Hei96: \citet{heintz96}; 
      Hen18: \citet{henry18}; 
      Irw92: \citet{irwin92};
      Irw96: \citet{irwin96};
      Izm19: \citet{izmailov19};
      Kna20: \citet{knapp20};     
      Leg19: \citet{leggett19};
      Liu12: \citet{liu12};
      Mam13: \citet{mamajek13};
      Man19: \citet{mann19};
      Mas21: \citet{mason21a};
      Mas23: \citet{mason23};
      Mon06: \citet{montagnier06};
      MSC:   Multiple Star Catalogue \citep{tokovinin97,tokovinin18};
      Nid02: \citet{nidever02};
      Pin18: \citet{pinamonti18};
      Pop13: \citet{pope13};
      Pou00: \citet{pourbaix00};
      Rob16: \citet{roberts16};
      Seg00: \citet{segransan00};
      Sha17: \citet{shakht17};
      Tok19: \citet{tokovinin19b};
      Tok21: \citet{tokovinin21};  
      Tok23: \citet{tokovinin23};
      Tok24: \citet{tokovinin24};
      Tok25: \citet{tokovinin25};
      Tsv24: \citet{tsvetkova24};
      Vri26: \citet{vrijmoet26};
      Wri13: \citet{wright13};
      Xua24: \citet{xuan24}.
     } 
   \footnotesize{$^{\text{(f)}}$ In italics, data calculated from $a$.}
   \footnotesize{$^{\text{(g)}}$ In italics, data calculated from $\alpha$.} 
   \footnotesize{$^{\text{(h)}}$ WDS 14164+1348: Mean separation derived from multiple measurements on a UKIDSS \citep{lawrence07} LAS J2 image of both stars using the manual `Measurer' tool in Aladin sky atlas \citep{bonnarel00}.}
 \end{justify}     
\end{landscape}

\bsp	
\label{lastpage}
    
\end{document}